\documentclass[aps,prb,twocolumn,groupedaddress]{revtex4-2}
\usepackage{graphicx,color}
\usepackage{dcolumn}
\usepackage{amsmath}
\usepackage{amssymb}
\usepackage{amsbsy}
\usepackage{graphics}
\usepackage{setspace}
\usepackage{array}
\usepackage{color}
\usepackage{fontenc}
\usepackage{textcomp}
\usepackage{rotating}
\usepackage{bm}
\usepackage{cancel}

\let\a=\alpha \let\b=\beta   

 \let\t=\tau

\def\nn{\nonumber}

\def\bpm{\begin{pmatrix}}
\def\epm{\end{pmatrix}}
\def\be{\begin{equation}}
\def\ee{\end{equation}}
\def\bea{\begin{eqnarray}}
\def\eea{\end{eqnarray}}
\def\ba{\begin{array}}
\def\ea{\end{array}}

\newcommand{\mathsym}[1]{{}}
\newcommand{\unicode}[1]{{}}

\newcommand{\besb}{\begin{subequations}}
\newcommand{\eesb}{\end{subequations}}
\newcommand{\beal}{\begin{align}}
\newcommand{\eeal}{\end{align}}

\hfuzz=\maxdimen
\begin{document}
\title{Probing magnetic anisotropy in Kagome antiferromagnetic Mn$_3$Ge with torque magnetometry}
\author{Y.~S.~Liu$^{1,2,3}$}
\author{H.~Xiao$^{4}$}
\author{A.~B.~Yu$^{5}$}
\author{Y.~F.~Wu$^{1}$}
\author{K.~Manna$^{6,7}$}
\author{Claudia Felser$^{6}$}
\author{C.~M.~Schneider$^{2,3}$}
\author{Hong-Yi~Xie$^{1\dag}$}
\author{T.~Hu$^{1\star}$}

\affiliation{$^{1}$Beijing Academy of Quantum Information Sciences, Beijing, 100193, China}
\affiliation{$^{2}$Fakult{\"a}t f{\"u}r Physik, Universit{\"a}t Duisburg-Essen, D-47057 Duisburg, Germany}
\affiliation{$^{3}$Peter Gr{\"u}nberg Institute PGI-6, Forschungszentrum J{\"u}lich, D-52425 J{\"u}lich, Germany}
\affiliation{$^{4}$Center for High Pressure Science and Technology Advanced Research, Beijing, 100094, China}
\affiliation{$^{5}$State Key Laboratory of Functional Materials for Informatics, Shanghai Institute of Microsystem and Information Technology, Chinese Academy of Sciences, Shanghai 200050, China}
\affiliation{$^{6}$Max Planck Institute for Chemical Physics of Solids, Dresden, Germany}
\affiliation{$^{7}$Indian Institute of Technology- Delhi, Hauz Khas, New Delhi 110 016, India}
\date{\today}

\begin{abstract}
We investigate the magnetic symmetry of the topological antiferromagnetic material $\mathrm{Mn_3Ge}$ by using torque measurements.
Below the N\'eel temperature, detailed angle-dependent torque measurements were performed on $\mathrm{Mn_3Ge}$ single crystals in directions parallel and perpendicular to the Kagome basal plane. The out-of plane torque data exhibit $\pm\sin\theta$ and $\sin2\theta$ behaviors, of which the former results from the spontaneous ferromagnetism within the basal plane and the latter from the in- and out-of-plane susceptibility anisotropy.
The reversible component of the in-plane torque exhibits $\sin6\varphi$ behavior, revealing the six-fold symmetry of the in-plane magnetic free energy.
Moreover, we find that the free energy minima are pinned to the direction of spontaneous ferromagnetism, which correspond to the maxima of the irreversible component of the in-plane torque. We provide an effective spin model to describe the in-plane magnetic anisotropy.
Our results demonstrate that the ground state of $\mathrm{Mn_3Ge}$ is described by the coexistence of a strong six-fold antichiral order and a weak ferromagnetic order induced by second-order spin anisotropy.
\end{abstract}

\maketitle
\section{introduction}
The topological antiferromagnets, combining the characteristics of nontrivial topology and antiferromagnetism, arouse significant interests in modern spintronics~\cite{jungwirth2016antiferromagnetic,baltz2018antiferromagnetic}.
Antiferromagnetic materials are ideal platforms for spintronics applications because of their low susceptibility to external magnetic fields and vanishing stray field.
In addition, exceptional Berry curvatures of the electron bands in topological antiferromagnets can give rise to the intrinsic anomalous Hall effect~\cite{nagaosa2010anomalous}, as strong as that observed in  conventional ferromagnetic materials~\cite{kubler2014non,nakatsuji2015large,nayak2016large,kubler2018weyl}.
Recently, the Kagome-type noncollinear antiferromagnetic compounds $\mathrm{Mn_3X}$, including the hexagonal compounds with X=Ge, Sn, and Ga, and
the cubic compounds with X=Ir and Pt, have attracted intensive studies~\cite{zhang2017strong,mukherjee2020sign}.
In particular, $\mathrm{Mn_3Ge}$ and $\mathrm{Mn_3Sn}$ were proposed to be antiferromagnetic Weyl semimetals~\cite{kuroda2017evidence,yang2017topological,kubler2018weyl,kubler2014non,chen2021anomalous}. A variety of intriguing phenomena were observed, such as large anomalous Hall effect~\cite{nayak2016large,nakatsuji2015large,iwaki2020large,mukherjee2020sign,liu2017transition,surgers2018electrical,liu2018electrical},
spin Hall effect~\cite{kimata2019magnetic,zhang2016giant,zhang2017strong}, anomalous Nernst effect~\cite{ikhlas2017large,hong2020large,wuttke2019berry,xu2020finite}, and magneto-optical Kerr effect~\cite{wu2020magneto}. Imaging and writing magnetic domains via the anomalous Nernst effect has also been realized~\cite{reichlova2019imaging}.

Among those compounds, $\mathrm{Mn_3Ge}$ has been found to exhibit a large intrinsic anomalous Hall effect even at room temperature and this effect is strongly anisotropic and can be switched with a small applied magnetic field~\cite{kiyohara2016giant,nayak2016large}.
These interesting transport properties are attributed to the peculiar magnetic patterns of the Mn spins.
The hexagonal Mn$_3$Ge is a noncollinear antiferromagnetic crystal with antichiral spin structure in the basal plane below the N\'eel temperature $T_\mathrm{N}$ ranging from $365 \, \mathrm{K}$ to $400\, \mathrm{K}$~\cite{ohoyama1961x,nagamiya1982triangular,tomiyoshi1983triangular,yamada1988magnetic,brown1990determination,qian2014exchange,nayak2016large}.
The resulting electron bands can support Weyl points that act as effective monopoles generating large Berry curvatures at the Fermi surface~\cite{kubler2014non,fang2003anomalous,vsmejkal2017route}.
Nevertheless, the magnetic anisotropies of the Mn$_3$X family have long been investigated and debated. The torque measurements in Mn$_3$Sn indicate a six-fold symmetry and another two-fold symmetry of the magnetic anisotropy~\cite{duan2015magnetic}. While a near cubic symmetry of the anisotropy is found in noncollinear antiferromagnet $\mathrm{L1_2}$-ordered Mn$_3$Ir, which is never found in ferromagnets~\cite{jenkins2019magnetic,vallejo2007measurement,chen2020manipulating}. In addition, an extremely strong second-order anisotropy appears in Mn$_3$Ir due to its frustrated triangular magnetic ground state~\cite{szunyogh2009giant}. For Mn$_3$Ge it was argued that no in-plane anisotropy energy exists up to four-fold terms~\cite{tomiyoshi1983triangular,nagamiya1982triangular,kiyohara2016giant},
because the second- and the fourth-order anisotropy energies from each Mn site cancel out and cannot induce significant energy landscapes~\cite{cable1993magnetic}.
However, a weak in-plane ferromagnetic moment with nearly isotropic susceptibility is detected in Mn$_3$Ge~\cite{yamada1988magnetic,tomiyoshi1983triangular}. The origin of this small ferromagnetism is not yet clarified and can be determined by either spin-orbit coupling effects~\cite{manna2018heusler,kubler2014non,nyari2019weak} or the local spin anisotropy~\cite{cable1993magnetic,liu2017anomalous,soh2020ground}. Besides, the magnetic anisotropy for fields applied along different crystallographic directions have not been thoroughly evaluated, despite of the availability of neutron diffraction studies~\cite{chen2020antichiral,soh2020ground}.

In this work, we use torque magnetometry as a sensitive probe to investigate the macroscopic magnetic anisotropy in hexagonal Mn$_3$Ge single crystals and provide a theoretical analysis based on effective Hamiltonian expressed in terms of magnetic order parameters. Torque magnetometry is a highly sensitive tool to detect the magnetic anisotropy of materials.
A finite torque, defined as $\boldsymbol{\tau}=\mathbf{m} \times\mathbf{H}$, arises as if the magnetization $\mathbf{m}$ is no longer collinear with the applied magnetic field $\mathbf{H}$.
For a fixed rotation axis, the reversible component of the torque is the rotation-angle derivative of the thermodynamic free energy, i.e., $\tau(\theta) = -dF/d\theta$, while the irreversible component of the torque can be induced by the free energy minima acting as intrinsic pinning centers~\cite{xiao2006angular,hu2012superconductivity}. In Sec.~\ref{exp}, we describe the experimental details. In Sec.~\ref{theory}, we introduce an effective spin model to analyze the observed magnetic anisotropy. In Sec.~\ref{conc} we summarize the results.

\begin{figure*}
	\includegraphics[trim=0cm 0cm 0cm 0cm, clip=true, width=0.6\textwidth]{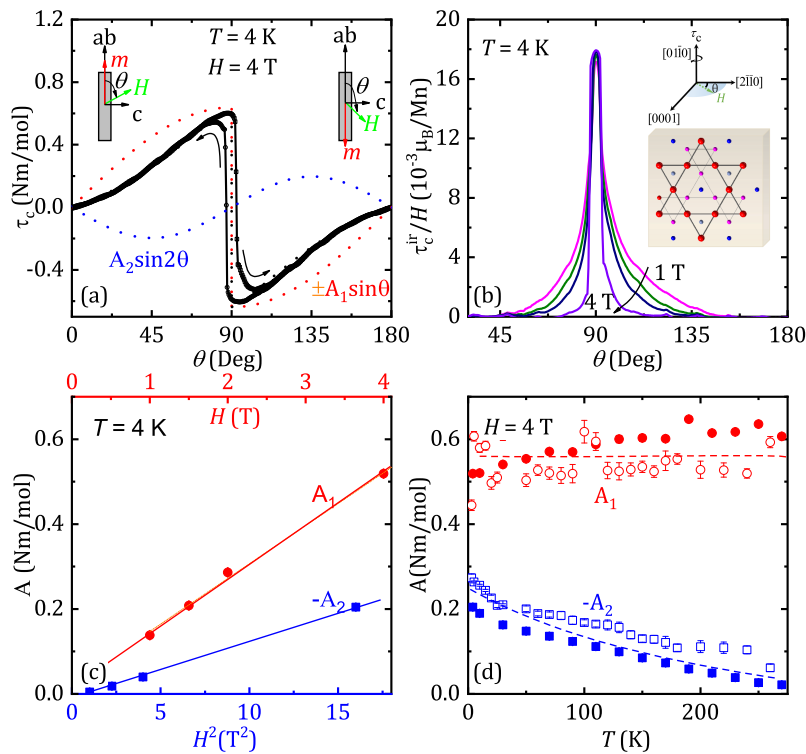}
	\caption{Out-of plane torque $\tau_{c}(\theta)$ measurements. The experimental geometry is depicted in the inset of panel (b).
		(a) $\tau_c$ as a function of $\theta$ sweeping in anticlockwise (right arrow) and clockwise (left arrow) directions at $T=4\,\mathrm{K}$ and $H=4\,\mathrm{T}$.
		    As defined in Eq.~\eqref{rev-t}, the reversible torque $\tau_c^{\mathrm{re}}(\theta)$ consists of two components $A_{1}$ (blue dashed line) and $A_2$ (red dashed line).
		(b) Normalized irreversible torque $\tau_c^{\mathrm{ir}}/H$ as a function of $\theta$ for magnetic fields $1\,\mathrm{T} \le H \le 4\,\mathrm{T}$ at $T=4\,\mathrm{K}$.
		(c) Coefficients $A_1$ (top axis and red symbols) and $A_2$ (bottom axis and blue symbols) as functions of $H$ and $H^2$, respectively, at $T=4\,\mathrm{K}$.
		(d) Coefficients $A_{1}$ (red symbols) and $A_2$ (blue symbols) as functions of temperature $T$ at $H=4\, \mathrm{T}$. The dash lines are guides to the eyes and the solid and hollow symbols are the results for the samples with the masses of 1.45 mg and 0.0157 mg, respectively.
	}
	\label{fig:Figure 1}
\end{figure*}

\section{Experiment} \label{exp}
\begin{figure*}
	\includegraphics[trim=0cm 0cm 0cm 0cm, clip=true, width=0.9\textwidth]{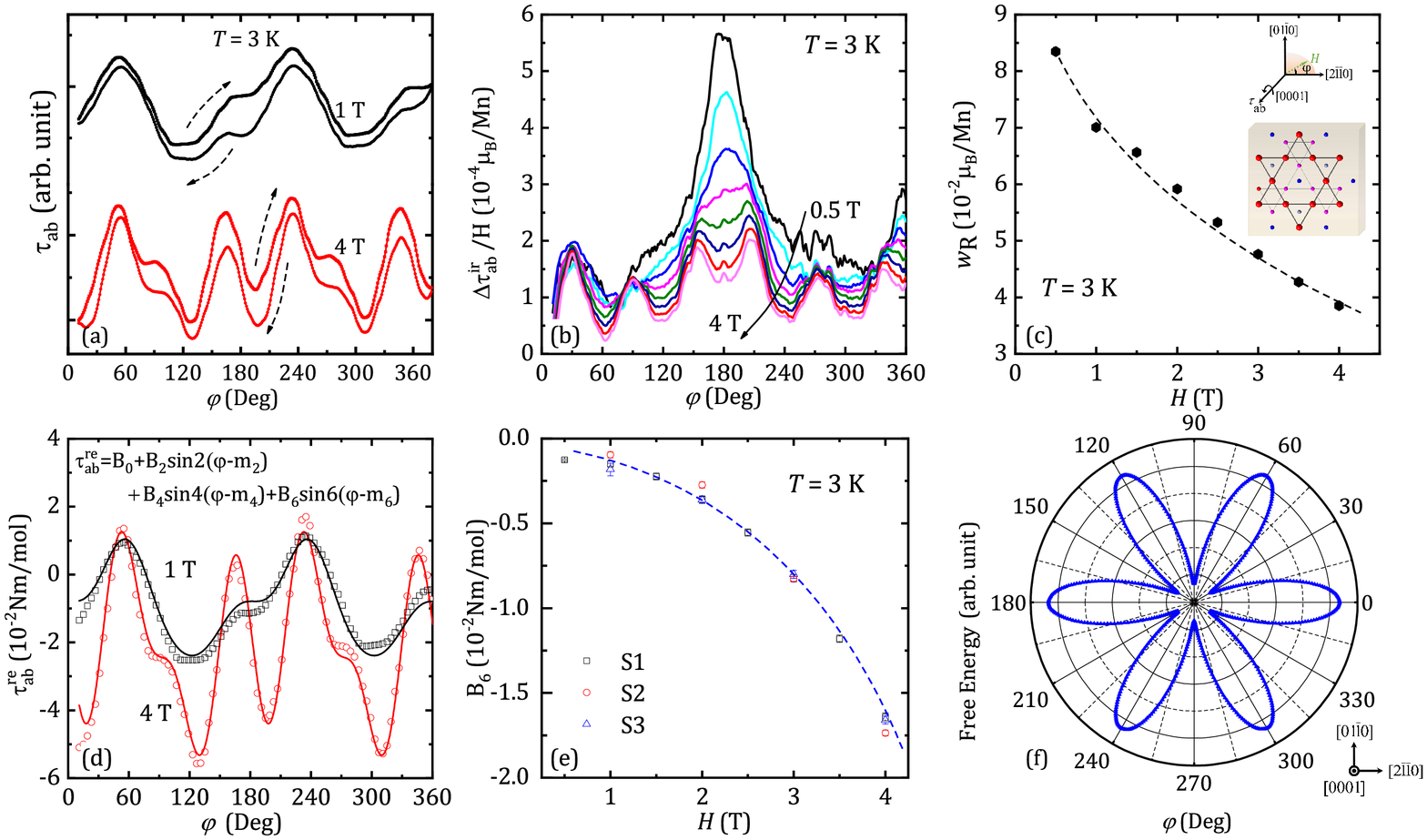}
	\caption{In-plane torque $\tau_{ab}(\varphi)$ measurements. The experimental geometry is depicted in the inset of panel (c).
		(a) $\tau_{ab}(\varphi)$ as a function of $\varphi$ sweeping in anticlockwise (right arrows) and clockwise (left arrows) directions at $T=3\,\mathrm{K}$ and $H=1\,\mathrm{T}$ (black lines) and $H=4 \,\mathrm{T}$ (red lines).
		(b) Normalized irreversible torque $\tau_{ab}^{\mathrm{ir}}/H$ as a function of $\varphi$ for $0.5 \,\mathrm{T} \le H \le 4\,\mathrm{T}$ T with a step of $0.5\,\mathrm{T}$ at $T=3 \,\mathrm{K}$.
		(c) Rotation hysteresis $W_\mathrm{R}$ as a function of $H$.
		(d) Reversible torque $\tau_{ab}^{\mathrm{re}}(\varphi)$ for $H= 1 \, \mathrm{T}$ (black square) and $H= 4 \, \mathrm{T}$ (red circle) at $T=3$ K. The solid lines are fitting results by using Eq.~\eqref{tau-ab}.
		(e) Coefficient $B_6$ as a function of $H$ at $T=4\,\mathrm{K}$. $\mathrm{S_1}$ and $\mathrm{S_2}$ indicate regular-shape samples of square cross section with masses of 5.9 mg and 3.1 mg, respectively, and $\mathrm{S_3}$ indicates an irregular-shape sample with mass of 4.5mg. The dashed lines are guides to the eyes.
		(f) Polar plot of the magnetic anisotropic free energy $F(\varphi)$ (blue line) obtained from the $B_6$ component of the reversible torque in (d) and the orientation of Mn$_3$Ge sample (red line).
	}
	\label{fig:Figure 2}
\end{figure*}

High-quality Mn$_3$Ge single crystals were synthesized by melting stoichiometric quantities of Mn and Ge elements using the Bridgman-Stockbarger technique~\cite{nayak2016large}.
The angle-dependent magnetic torques of Mn$_3$Ge were performed by using a piezoresistive torque magnetometer in the Quantum Design Physical Property Measurement System. The experimental geometries for out-of plane and in-plane rotations are shown in the insets of Figs.~\ref{fig:Figure 1}(b) and \ref{fig:Figure 2}(c), respectively.

Figure~\ref{fig:Figure 1}(a) shows the out-of-plane angle $\theta$ dependence of torque ($\tau_c$) in anticlockwise ($\tau_{c}^{+}$) and clockwise ($\tau_{c}^{-}$) directions at $T=4\,\mathrm{K}$ and $H=4\,\mathrm{T}$.
It is found that, when the magnetic field is nearly along the $c$ axis ($\theta \approx 90^\circ$), $\tau_c(\theta)$ switches to the opposite value and exhibits a large torque hysteresis loop in one angle-sweep circle. Such a hysteresis loop indicates magnetic pinning of a spontaneous ferromagnetic moment $\mathbf{m}$ within the basal plane.
As discussed in Sec.~\ref{theory}, this ferromoment is induced by an in-plane noncollinear antiferromagnetic (antichiral) order.
It is evident that the basal plane is the energetically most favored plane of spontaneous magnetic symmetry breaking dominated by the Dzyaloshinskii-Moriya interaction~\cite{qian2014exchange,yamada1988magnetic,tomiyoshi1982magnetic,nagamiya1982triangular,soh2020ground,chen2020antichiral}.
As sketched in the inset of Fig.~\ref{fig:Figure 1}(a), the ferromoment $\mathbf{m}$ flips within the basal plane as the in-plane component of the applied field goes to the opposite direction, which, therefore, results in a sign reverse of the torque about $\theta=90^\circ$.
As shown in the inset of Fig.~\ref{fig:Figure 1}(b), we find that $\mathbf{m}$ is parallel to the $[2\bar{1}\bar{1}0]$ direction.
Moreover, we estimate its amplitude $m$ based on the irreversible torque $\tau_c^{\mathrm{ir}} \equiv \tau_{c}^{+}-\tau_{c}^{-}=2 m H$ at $\theta=90^\circ$.
Figure~\ref{fig:Figure 1}(b) shows the normalized irreversible torque $\tau_c^{\mathrm{ir}}(\theta)/H$ as a function of $\theta$ measured for $1 \, \mathrm{T} \leq H \leq 4\,\mathrm{T}$.
We find that $\tau_c^{\mathrm{ir}}(\theta=90^\circ)/H= 0.018\,\mu_\mathrm{B}$ and the value is field-independent. This gives $m=0.009 \, \mu_\mathrm{B}/\mathrm{Mn}$, which is consistent with the spontaneous in-plane magnetization $0.007 \, \mu_\mathrm{B}/\mathrm{Mn}$ observed in various studies~\cite{yamada1988magnetic,nagamiya1982triangular,nayak2016large}. As shown in Eq.~\eqref{A-M}, this small ferromoment reflects the coupling between the antichiral order [Fig.~\ref{OP}(c)] and the ferromagnetic order [Fig.~\ref{OP}(d)] via second-order spin anisotropy energy.

The spontaneous ferromoment is also observed in the reversible component of the torque, i.e., $\tau_c^{\mathrm{re}} \equiv (\tau_{c}^{+}+\tau_{c}^{-})/2$. In Fig.~\ref{fig:Figure 1}(a), we fit $\tau_c^{\mathrm{re}} (\theta)$ data by the formula
\be \label{rev-t}
\tau_c^{\mathrm{re}} (\theta) = \pm A_1\sin\theta+A_2\sin2\theta.
\ee
Here, the $A_1$ component (the red dotted curve) results from the spontaneous ferromoment, which takes the positive sign ``$+$'' for $0<\theta<\pi/2$ and the negative sign ``$-$''  for $\pi/2<\theta<\pi$, so that it changes suddenly at $\theta=\pi/2$. The $A_2$ component (the blue dotted curve) originates from the difference between the in-plane and out-of-plane susceptibilities. We show the field and temperature dependence of $A_{1,2}$ in Figs.~\ref{fig:Figure 1}(c) and ~\ref{fig:Figure 1}(d).

In Fig.~\ref{fig:Figure 1}(c), we observe that $A_1$ increases linearly with $H$, which confirms a constant in-plane magnetic moment $m=A_1/H=0.124\pm 0.005 \, \mathrm{Nm/mol/T}$ $(0.0075\pm 0.0003 \, \mu_\mathrm{B}/\mathrm{Mn})$, constant with that found in the irreversible torque in Fig.~\ref{fig:Figure 1}(a). As shown in Fig.~\ref{fig:Figure 1}(d), $m$ is temperature-independent up to $T \approx 275\,\mathrm{K}$, which is still far below the N\'eel temperature. This evidences the intrinsic correlation of the ferromagnetic and antiferromagnetic orders~\cite{nayak2016large,yamada1988magnetic,tomiyoshi1983triangular,nagamiya1982triangular}, as discussed in Sec.~\ref{theory}.
On the other hand, from the $A_2$ component we can obtain the susceptibility anisotropy $\Delta\chi \equiv \chi_{ab}-\chi_{c}$, where $\chi_{c}$ and $\chi_{ab}$ are the out-of- and in-plane susceptibilities, respectively, because the field-induced magnetization produces a torque equal to
$\frac{1}{2}\Delta\chi  H^{2}\sin{2\theta}$~\cite{herak2015torque,hong2016large}. In Fig.~\ref{fig:Figure 1}(c), we show that $A_2$ is negatively proportional to $H^2$ and obtain $\Delta\chi =-0.0278 \mathrm{m^3/mol}$  at $T=4\,\mathrm{K}$, which is consistent with the previous studies~\cite{yamada1988magnetic,nayak2016large}. Figure~\ref{fig:Figure 1}(d) indicates that $\Delta\chi <0$ in the temperature range below $275\,\mathrm{K}$ and tends to go to zero as the temperature increases.
The low-temperature susceptibility anisotropy reflects the strength of the Dzyaloshinskii-Moriya interaction~\cite{liu2017anomalous}, while the high-temperature behavior $\chi_{ab} \to \chi_{c}$ ought to be the signal of suppression of magnetic anisotropy as the N\'eel temperature is approached from below.
It is worthy to note that the data in Fig.~\ref{fig:Figure 1}(d) are taken from the two distinct samples with masses of $1.45 \, \mathrm{mg}$ and $0.0157\,\mathrm{mg}$, where the $0.0157\,\mathrm{mg}$ is derived from the torque signal since it is far too small to measure on an ordinary balance. 

In Fig.~\ref{fig:Figure 2}(a), we show the in-plane angle $\varphi$ dependence of the torque $\tau_{ab}$ in anticlockwise and clockwise directions at $T=3\,\mathrm{K}$ and $H=1\,\mathrm{T}$ and $4\,\mathrm{T}$. We observe that $\tau_{ab}(\varphi)$ exhibits hysteresis at almost any angle, which signals magnetic pinning as discussed in Sec.~\ref{theory}.
Figure~\ref{fig:Figure 2}(b) shows the normalized irreversible torque $\tau_{ab}^{\mathrm{ir}}/H$ as a function of $\varphi$ for various field strengths ranging from $0.5\,\mathrm{T}$ to $4\,\mathrm{T}$.
We observe that $\tau_{ab}^{\mathrm{ir}}/H$ exhibits two peaks for $H< 1.5\,\mathrm{T}$ and six peaks at $\varphi_n=30^\circ+(n-1) \times 60^\circ$ ($1 \le n \le 6$) for $H>1.5\,\mathrm{T}$.
The six peaks originate from the six-fold in-plane magnetic anisotropy due to the crystal field effect [see Eq.~\eqref{decp} and Ref.~\onlinecite{fujita2017field}].
In Fig.~\ref{fig:Figure 2}(c), we show the rotational hysteresis, define by the area of the hysteresis loop $W_\mathrm{R} \equiv \int_{0}^{2\pi}\tau_{ab}^{\mathrm{ir}}(\varphi) \, d\varphi$ and giving twice the energy lost in rotating the magnetization, as a function of magnetic field.
The high-field rotational hysteresis reveals the existence of the unidirectional anisotropy due to the coupling between ferromagnetic and antiferromagnetic orders~\cite{meiklejohn1957new, nogues1999exchange}, which occurs in parallel with the exchange bias in the magnetization curves of Mn$_3$Ge~\cite{qian2014exchange}.

In Fig.~\ref{fig:Figure 2}(d), we fit the reversible torque $\tau_{ab}^{\mathrm{re}}(\varphi)$ using the formula
\begin{align}
\tau_{ab}^{\mathrm{re}}(\varphi)=
&\, B_0+B_2\sin [2(\varphi-m_2)] \nn \\
&\, +B_4\sin[4(\varphi-m_4)] +B_6\sin[6(\varphi-m_6)]. \label{tau-ab}
\end{align}
The fitting result of $B_6$ is shown in Fig.~\ref{fig:Figure 2}(e), where the black and red hollow symbols are for regular-shape samples of square cross section with masses of 5.9 mg and 3.1 mg, respectively, and the blue symbols for irregular-shape sample with mass of 4.5 mg.
We observe that $B_6$ is negative and sample-independent and its magnitude increases as $H$ increasing.
This observation suggests the intrinsic in-plane anisotropy properties of Mn$_3$Ge and rules out extrinsic effects such as magnetoelastic forces. We note that $B_6$ can be directly related to the six-fold anisotropy energy of the antichiral order as discussed in Sec.~\ref{theory}.  We note that the fitting result of $B_{0,2,4}$ is not included in Fig.~\ref{fig:Figure 2}(e), because these components are plausibly extrinsic: $B_0$ comes from the background of torque magnetometer, and $B_{2,4}$ are different in magnitude for the three samples and likely induced by the artificial magnetoelastic anisotropy, which cannot indicate the intrinsic magnetization with respect to the crystallographic axes~\cite{gomonay2005mechanism,gomonay2007shape}.

In order to further investigate the six-fold symmetry of the magnetic anisotropy in Mn$_3$Ge, we convert the reversible torque into magnetic free energy by the definition $F(\varphi) = - \int \tau_{ab}^{\mathrm{re}}(\varphi)\,d\varphi$ up to a constant.
In Fig.~\ref{fig:Figure 2}(f), we show the free energy (blue line) and the related crystal orientation of Mn$_3$Ge samples (red line) in the polar coordinates.
We find free energy minima $\varphi_n=30^\circ + (n-1)\times 60^\circ$ ($1 \le n\le 6$), which coincide with the peaks of the irreversible torques at high field as shown in Fig.~\ref{fig:Figure 2}(b). Not as a surprise, the symmetry of free energy respecting the Kagome lattice is consistent with the symmetry obtained by the neutron diffraction studies~\cite{soh2020ground,chen2020antichiral}. This confirms that the magnetic easy axis points aiming the $[01\bar{1}0]$  direction as shown in the insets of Figs.~\ref{fig:Figure 1}(b) and ~\ref{fig:Figure 2}(c). In the following section, we characterize the in-plane anisotropy by an effective spin Hamiltonian.

\section{In-plane anisotropic Hamiltonian} \label{theory}

\begin{figure}
\includegraphics[width=0.47\textwidth]{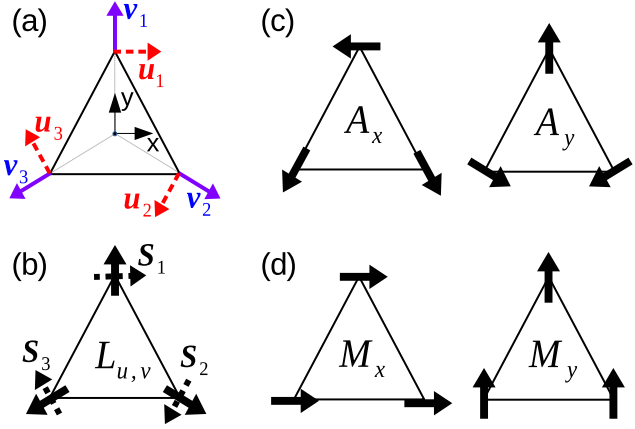}
\caption{Spin configurations corresponding to the order parameters in Eq.~\eqref{ops}.
(a) Local orthogonal easy axes $\mathbf{u}_{1,2,3}$ (red) and $\mathbf{v}_{1,2,3}$ (blue).
(b) Chiral orders $L_{u}$ (solid) and $L_{v}$ (dotted).
(c) Antichiral order $\mathbf{A}$.
(d) Ferromagnetic order $\mathbf{M}$. Experiment date in Figs.~\ref{fig:Figure 1}(a) and \ref{fig:Figure 2}(f) indicate that the ground state prefers the coexistence of $A_y$ and $M_y$.}
\label{OP}
\end{figure}

It has been suggested that the ground-state magnetic phase of $\mathrm{Mn_3 Ge}$ is described by two in-plane vector orders, i.e., an antichiral (AC) order [Fig.~\ref{OP}(c)] and a weak ferromagnetic (FM) order [Fig.~\ref{OP}(d)]~\cite{chen2020antichiral, soh2020ground}. The AC order reduces the magnetic symmetry from hexagonal $D_{6h}$ to rhombic $D_{2h}$ and the FM order furthermore reduces $D_{2h}$ to monoclinic $C_{2h}$~\cite{cable1993magnetic}. Assuming a uniform spin configuration respecting the translation and inversion symmetries of the $\mathrm{Mn_3 Ge}$ lattice, we consider the effective triple-spin JDC Hamiltonian describing the energy per unit cell~\cite{liu2017anomalous, soh2020ground,chen2020antichiral},
\begin{subequations} \label{ham-0}
\begin{align}
\mathcal{H} = &\, J \sum_{\langle jk \rangle} \mathbf{S}_j \cdot \mathbf{S}_k + D \sum_{\langle jk \rangle} \hat{\mathbf{z}} \cdot (\mathbf{S}_j \times \mathbf{S}_k) \nn \\
    &\, + \sum_{l=1}^{3}\sum_{m=-l}^{l} C_{2 l}^{2m} T_{2l}^{2m} - 2g\mu_\mathrm{B} \mathbf{H} \cdot \sum_{j} \mathbf{S}_j, \label{ham-1}
\end{align}
\end{subequations}
where $J> 0$ is the sum of the Heisenberg exchanges, which favors noncollinear antiferromagnetic orders [Figs.~\ref{OP}(b) and \ref{OP}(c)],
$D<0$ is the Dzyaloshinskii-Moriya (DM) interaction mediated by the spin-orbit coupling that aligns the spins in a $120^\circ$ structure in the $ab$ plane and favors the antichiral order [Fig.~\ref{OP}(c)], $C_{2 l}^{2m}$ represents the crystal field energy respecting the $C_{2h}$ symmetry, and $\mathbf{H}$ is the applied magnetic field with $g$ denoting the Land\'e $g$ factor.
In Eq.~\eqref{ham-1}, the crystal field tensors $T_{2l}^{2m}$ are defined by the Stevens operators $O_{2 l}^{2m}$~\cite{bauer2009mag}: $T_{2l}^{2m} \equiv 3^{2l-1}\sum_{j=1}^3 O_{2 l}^{2m}(\mathbf{S}_j)$ and $O_{2 l}^{0} (\mathbf{S}) = \mathbf{S}^{2l}$, $O_{2l}^{2l}(\mathbf{S}) = (S_{+}^{2l} + S_{-}^{2l})/2$, $O_{2l}^{-2l}(\mathbf{S}) = (S_{+}^{2l} - S_{-}^{2l})/(2 i)$, $O_{2l}^{\pm 2m} (\mathbf{S}) = \mathbf{S}^{2l-2m} O_{2m}^{\pm 2m}(\mathbf{S})$ with $1 \le m \le l-1$, where $S_{j,\pm} \equiv (\mathbf{u}_j \pm i \mathbf{v}_j) \cdot \mathbf{S}_j$ is defined with respect to the local easy axes $\mathbf{u}_j = (\cos[2\pi(j-1)/3], \,   \sin[2\pi(j-1)/3])$ and $\mathbf{v}_j = (\sin[2\pi(j-1)/3], \,  -\cos[2\pi(j-1)/3])$, as shown in Fig.~\ref{OP}(a). We parameterize the in-plane spins by six order parameters~\cite{soh2020ground}
\begin{align} \label{ops}
& L_u = \sum_{j=1}^3 \mathbf{u}_j \cdot \mathbf{S}_j, \quad L_v = \sum_{j=1}^3 \mathbf{v}_j \cdot \mathbf{S}_j, \quad
\mathbf{M} = \sum_{i=1}^3 \mathbf{S}_i, \nn \\
& A_x = -(\mathbf{u}_1 \cdot \mathbf{S}_1 + \mathbf{u}_2 \cdot \mathbf{S}_3 + \mathbf{u}_3 \cdot \mathbf{S}_2), \nn \\
& A_y = \mathbf{v}_1 \cdot \mathbf{S}_1 + \mathbf{v}_2 \cdot \mathbf{S}_3 + \mathbf{v}_3 \cdot \mathbf{S}_2,
\end{align}
where the chiral orders $L_{u}$, $L_{v}$ [Fig.~\ref{OP}(b)], and the AC order $\mathbf{A}$ [Figs.~\ref{OP}(c)] and the FM order $\mathbf{M}$ [Figs.~\ref{OP}(d)] are transformed according to the irreducible representations $B_{1g}$, $B_{2g}$, and $E_{1g}$ of the $D_{6h}$ group, respectively. We parameterize $\mathbf{M}=M(\cos\a, \sin\a)$, $\mathbf{A}=A(\cos\b, \sin\b)$, and $\mathbf{H}=H(\cos\varphi, \sin\varphi)$ and rewrite the Hamiltonian in terms of the order parameters,
\begin{align}
& H = a A^2 + b  A^4 + c_{6\b} A^6 + d M^2 + b M^4 + c_{6\a} M^6 \nn \\
& + \sum_{k=1}^{3} \sum_{l=1}^{2k-1} f_{2k-l,l}(\a,\b)A^{2k-l} M^l - 2 g \mu_\mathrm{B} H M \cos(\varphi-\a), \label{decp}
\end{align}
with the coefficients
\begin{align}
& a = -J/6, \,\, b =  C_4^0, \,\, d=J/3, \,\, c_{6\b} = C_{6}^0+C_6^6 \cos 6 \b, \nn \\
& f_{1,1}(\a,\b)= -2 C_{2}^2 \cos(\a-\b), \nn \\
& f_{1,3}(\a,\b)= -3C_{4}^2 \cos(\a-\b), \nn \\
& f_{1,5}(\a,\b)= -4 C_{6}^2 \cos(\a-\b)-C_{6}^4 \cos(5 \a+\b), \nn \\
& f_{2,2}(\a,\b)= 4 C_{4}^0 + 6 C_4^4 \cos(2\a-2\b), \nn \\
& f_{2,4}(\a,\b)= 9 C_6^0 + C_6^2 \cos(4\a+2\b) + 10 C_6^4 \cos(2\a-2\b), \nn \\
& f_{3,3}(\a,\b) = -2C_6^0 \cos(3\a+3\b) - 12 C_6^2 \cos(\a-\b) \nn \\
& \hspace{1.8cm} - 20 C_6^6 \cos(3\a-3\b), \label{coef}
\end{align}
and $f_{l,m}(\a,\b)=f_{m,l}(\b,\a)$, where we have taken $L_{u,v}=0$ and applied the energy hierarchy $J \gg |D| \gg |C_{l}^{m}| \gg |C_{l}^{-m}|$ for $m > 0$ according to experimental observations~\cite{chen2020antichiral, soh2020ground}. Furthermore, our experiment suggests that it is reasonable to neglect the fourth order term $C_4^0=0$ and keep only the lowest-order AC-FM coupling $C_2^2$. The Hamiltonian \eqref{decp} is simplified to
\begin{align} \label{ener}
H = &\, a A^2 + c_{6\b} A^6 + d M^2 + c_{6\a} M^6
    -2 C_2^2 A M \cos(\a-\b) \nn \\
    &\, - 2 g \mu_\mathrm{B} H M \cos(\varphi-\a).
\end{align}
Minimizing the energy~\eqref{ener} with respect to $(M,A,\a,\b)$, we obtain the ground state as well as the metastable states. First, we note that the spontaneous FM order vanishes in the absence of AC-FM coupling, because $d >0$. We conclude that
\be
C_6^0 > C_{6}^6 > 0, \quad C_{2}^2 >0,
\ee
leading to the ground state $\a=\b=\pi/2$, that is, the configurations $M_y$ and $A_y$ in Figs.~\ref{OP}(c) and \ref{OP}(d), coinciding with the observations in Figs.~\ref{fig:Figure 1} and \ref{fig:Figure 2} and in Refs.~\cite{soh2020ground} and \cite{chen2020antichiral}.
The amplitudes of the ground-state order parameters read
\be  \label{A-M}
M= \frac{3 C_2^2A}{J}, \quad A =  \left(\frac{J}{18 \Delta_6}\right)^{1/4},
\ee
where $\Delta_6 = C_{6}^0-C_6^6 >0$. Therefore, the AC-induce FM order is weak $M/A \sim C_2^2 / J \ll 1$. In the presence of the magnetic field, we obtain the in-plane torque $\t_{ab} = 3 C_6^6 (A^6 \sin6\b + M^6\sin 6\a)$ where $(A,M,\a,\b)$ can take the values in ground state or any metastable state. In high-field limit $g \mu_B H \gg C_6^6 (A^6+M^6)$, we can estimate the reversible torque by replacing the angles $\a$ and $\b$ by $\varphi$ and obtain $\t_\mathrm{ab}^\mathrm{re} \approx 3 C_6^6 (A^6+M^6) \sin(6\varphi)$, which coincide with the symmetry of the reversible torque in Figs.~\ref{fig:Figure 2}(e) and \ref{fig:Figure 2}(f). In addition, the torque rotational hysteresis $W_\mathrm{R}$ in Fig.~~\ref{fig:Figure 2}(c) is attributed to the interplay of the six-fold anisotropy $C_6^6$ and the AC-FM coupling $C_2^2$~\cite{meiklejohn1957new, nogues1999exchange}.
Combing our experimental data and Eq.~\eqref{A-M}, we estimate the equilibrium orders $M = 3\times 0.0075/g$ and $A= 3\times 2.2/g$, where the factor ``3'' captures the number of spins. Further exploiting the susceptibility anisotropy expression $\Delta \chi = 8\sqrt{3} g^2 \mu_\mathrm{B}^2 D/(3 J^2)$ and taking $g=2$ and the value of $J$ in Ref.~\onlinecite{liu2017anomalous}, we estimate the relevant energy parameters in Eq.~\eqref{ener}, $J=22.4\,\mathrm{meV}$, $D= -2.54 \, \mathrm{meV}$, $C_2^2 = 2.5 \times 10^{-2} \, \mathrm{meV}$, $\Delta_6 = 1.0 \times 10^{-2} \, \mathrm{meV}$. Estimating the value of $C_6^6$ relies on a thorough numerical analysis of the model in Eq.~\eqref{decp}, which deserves a separated theoretical work.

\section{Summary} \label{conc}
In summary, we investigated the anisotropic magnetic properties of the antiferromagnet Mn$_3$Ge by torque magnetometry.
A spontaneous ferromoment of $m=(7.5\pm 0.3) \times 10^{-3} \, \mu_\mathrm{B}/\mathrm{Mn}$ is found to arise within the Kagome basal plane in $[01\bar{1}0]$ direction [Fig.~\ref{fig:Figure 1}(b)]. Furthermore, in-plane magnetic free energy possesses a six fold symmetry with the minima at $\varphi_n=30^\circ + (n-1)\times 60^\circ$ ($1 \le n\le 6$). These minima weakly pin the spontaneous ferromoment and cause the maxima of irreversible in-plane torque. Exploiting an effective Hamiltonian we estimated the values of various coupling energies, and demonstrate that the ground state of Mn$_3$Ge is characterized by a strong AC order ``ferromagnetically'' coupling to a weak AC-order-induced FM order with unidirectional anisotropy along $[01\bar{1}0]$ axis.
The two- and four-fold in-plane symmetries cannot been determined.


\begin{acknowledgments}
T.H. acknowledge the support of NSFC Grant No. 11574338. The work of H.-Y.X. was supported by NSFC under Grant No.~12074039.
H.X. acknowledge the support of NSAF Grant No. U1530402. K.M. and C.F. acknowledges financial support by European Research Council (ERC) Advanced Grant No. 742068 (“TOPMAT”), Deutsche Forschungsgemeinschaft (DFG) under SFB 1143 (Project No. 247310070) and W{\"u}rzburg-Dresden Cluster of Excellence on Complexity and Topology in Quantum Matter-ct.qmat (EXC 2147, project no. 39085490). K.M. acknowledges Max Plank Society for the funding support under Max Plank–India partner group project and Board of Research in Nuclear Sciences (BRNS) under 58/20/03/2021-BRNS/37084/ DAE-YSRA.
\end{acknowledgments}

$^{*}$ hutao@baqis.ac.cn
$^{\dag}$ xiehy@baqis.ac.cn

\begin{thebibliography}{55}%
\makeatletter
\providecommand \@ifxundefined [1]{%
 \@ifx{#1\undefined}
}%
\providecommand \@ifnum [1]{%
 \ifnum #1\expandafter \@firstoftwo
 \else \expandafter \@secondoftwo
 \fi
}%
\providecommand \@ifx [1]{%
 \ifx #1\expandafter \@firstoftwo
 \else \expandafter \@secondoftwo
 \fi
}%
\providecommand \natexlab [1]{#1}%
\providecommand \enquote  [1]{``#1''}%
\providecommand \bibnamefont  [1]{#1}%
\providecommand \bibfnamefont [1]{#1}%
\providecommand \citenamefont [1]{#1}%
\providecommand \href@noop [0]{\@secondoftwo}%
\providecommand \href [0]{\begingroup \@sanitize@url \@href}%
\providecommand \@href[1]{\@@startlink{#1}\@@href}%
\providecommand \@@href[1]{\endgroup#1\@@endlink}%
\providecommand \@sanitize@url [0]{\catcode `\\12\catcode `\$12\catcode
  `\&12\catcode `\#12\catcode `\^12\catcode `\_12\catcode `\%12\relax}%
\providecommand \@@startlink[1]{}%
\providecommand \@@endlink[0]{}%
\providecommand \url  [0]{\begingroup\@sanitize@url \@url }%
\providecommand \@url [1]{\endgroup\@href {#1}{\urlprefix }}%
\providecommand \urlprefix  [0]{URL }%
\providecommand \Eprint [0]{\href }%
\providecommand \doibase [0]{https://doi.org/}%
\providecommand \selectlanguage [0]{\@gobble}%
\providecommand \bibinfo  [0]{\@secondoftwo}%
\providecommand \bibfield  [0]{\@secondoftwo}%
\providecommand \translation [1]{[#1]}%
\providecommand \BibitemOpen [0]{}%
\providecommand \bibitemStop [0]{}%
\providecommand \bibitemNoStop [0]{.\EOS\space}%
\providecommand \EOS [0]{\spacefactor3000\relax}%
\providecommand \BibitemShut  [1]{\csname bibitem#1\endcsname}%
\let\auto@bib@innerbib\@empty
\bibitem [{\citenamefont {Jungwirth}\ \emph {et~al.}(2016)\citenamefont
  {Jungwirth}, \citenamefont {Marti}, \citenamefont {Wadley},\ and\
  \citenamefont {Wunderlich}}]{jungwirth2016antiferromagnetic}%
  \BibitemOpen
  \bibfield  {author} {\bibinfo {author} {\bibfnamefont {T.}~\bibnamefont
  {Jungwirth}}, \bibinfo {author} {\bibfnamefont {X.}~\bibnamefont {Marti}},
  \bibinfo {author} {\bibfnamefont {P.}~\bibnamefont {Wadley}},\ and\ \bibinfo
  {author} {\bibfnamefont {J.}~\bibnamefont {Wunderlich}},\ }\href@noop {}
  {\bibfield  {journal} {\bibinfo  {journal} {Nature Nanotechnology}\ }\textbf
  {\bibinfo {volume} {11}},\ \bibinfo {pages} {231} (\bibinfo {year}
  {2016})}\BibitemShut {NoStop}%
\bibitem [{\citenamefont {Baltz}\ \emph {et~al.}(2018)\citenamefont {Baltz},
  \citenamefont {Manchon}, \citenamefont {Tsoi}, \citenamefont {Moriyama},
  \citenamefont {Ono},\ and\ \citenamefont
  {Tserkovnyak}}]{baltz2018antiferromagnetic}%
  \BibitemOpen
  \bibfield  {author} {\bibinfo {author} {\bibfnamefont {V.}~\bibnamefont
  {Baltz}}, \bibinfo {author} {\bibfnamefont {A.}~\bibnamefont {Manchon}},
  \bibinfo {author} {\bibfnamefont {M.}~\bibnamefont {Tsoi}}, \bibinfo {author}
  {\bibfnamefont {T.}~\bibnamefont {Moriyama}}, \bibinfo {author}
  {\bibfnamefont {T.}~\bibnamefont {Ono}},\ and\ \bibinfo {author}
  {\bibfnamefont {Y.}~\bibnamefont {Tserkovnyak}},\ }\href@noop {} {\bibfield
  {journal} {\bibinfo  {journal} {Reviews of Modern Physics}\ }\textbf
  {\bibinfo {volume} {90}},\ \bibinfo {pages} {015005} (\bibinfo {year}
  {2018})}\BibitemShut {NoStop}%
\bibitem [{\citenamefont {Nagaosa}\ \emph {et~al.}(2010)\citenamefont
  {Nagaosa}, \citenamefont {Sinova}, \citenamefont {Onoda}, \citenamefont
  {MacDonald},\ and\ \citenamefont {Ong}}]{nagaosa2010anomalous}%
  \BibitemOpen
  \bibfield  {author} {\bibinfo {author} {\bibfnamefont {N.}~\bibnamefont
  {Nagaosa}}, \bibinfo {author} {\bibfnamefont {J.}~\bibnamefont {Sinova}},
  \bibinfo {author} {\bibfnamefont {S.}~\bibnamefont {Onoda}}, \bibinfo
  {author} {\bibfnamefont {A.~H.}\ \bibnamefont {MacDonald}},\ and\ \bibinfo
  {author} {\bibfnamefont {N.~P.}\ \bibnamefont {Ong}},\ }\href@noop {}
  {\bibfield  {journal} {\bibinfo  {journal} {Reviews of modern physics}\
  }\textbf {\bibinfo {volume} {82}},\ \bibinfo {pages} {1539} (\bibinfo {year}
  {2010})}\BibitemShut {NoStop}%
\bibitem [{\citenamefont {K{\"u}bler}\ and\ \citenamefont
  {Felser}(2014)}]{kubler2014non}%
  \BibitemOpen
  \bibfield  {author} {\bibinfo {author} {\bibfnamefont {J.}~\bibnamefont
  {K{\"u}bler}}\ and\ \bibinfo {author} {\bibfnamefont {C.}~\bibnamefont
  {Felser}},\ }\href@noop {} {\bibfield  {journal} {\bibinfo  {journal} {EPL
  (Europhysics Letters)}\ }\textbf {\bibinfo {volume} {108}},\ \bibinfo {pages}
  {67001} (\bibinfo {year} {2014})}\BibitemShut {NoStop}%
\bibitem [{\citenamefont {Nakatsuji}\ \emph {et~al.}(2015)\citenamefont
  {Nakatsuji}, \citenamefont {Kiyohara},\ and\ \citenamefont
  {Higo}}]{nakatsuji2015large}%
  \BibitemOpen
  \bibfield  {author} {\bibinfo {author} {\bibfnamefont {S.}~\bibnamefont
  {Nakatsuji}}, \bibinfo {author} {\bibfnamefont {N.}~\bibnamefont
  {Kiyohara}},\ and\ \bibinfo {author} {\bibfnamefont {T.}~\bibnamefont
  {Higo}},\ }\href@noop {} {\bibfield  {journal} {\bibinfo  {journal} {Nature}\
  }\textbf {\bibinfo {volume} {527}},\ \bibinfo {pages} {212} (\bibinfo {year}
  {2015})}\BibitemShut {NoStop}%
\bibitem [{\citenamefont {Nayak}\ \emph {et~al.}(2016)\citenamefont {Nayak},
  \citenamefont {Fischer}, \citenamefont {Sun}, \citenamefont {Yan},
  \citenamefont {Karel}, \citenamefont {Komarek}, \citenamefont {Shekhar},
  \citenamefont {Kumar}, \citenamefont {Schnelle}, \citenamefont {K{\"u}bler},
  \citenamefont {Felser},\ and\ \citenamefont {Parkin}}]{nayak2016large}%
  \BibitemOpen
  \bibfield  {author} {\bibinfo {author} {\bibfnamefont {A.~K.}\ \bibnamefont
  {Nayak}}, \bibinfo {author} {\bibfnamefont {J.~E.}\ \bibnamefont {Fischer}},
  \bibinfo {author} {\bibfnamefont {Y.}~\bibnamefont {Sun}}, \bibinfo {author}
  {\bibfnamefont {B.}~\bibnamefont {Yan}}, \bibinfo {author} {\bibfnamefont
  {J.}~\bibnamefont {Karel}}, \bibinfo {author} {\bibfnamefont {A.~C.}\
  \bibnamefont {Komarek}}, \bibinfo {author} {\bibfnamefont {C.}~\bibnamefont
  {Shekhar}}, \bibinfo {author} {\bibfnamefont {N.}~\bibnamefont {Kumar}},
  \bibinfo {author} {\bibfnamefont {W.}~\bibnamefont {Schnelle}}, \bibinfo
  {author} {\bibfnamefont {J.}~\bibnamefont {K{\"u}bler}}, \bibinfo {author}
  {\bibfnamefont {C.}~\bibnamefont {Felser}},\ and\ \bibinfo {author}
  {\bibfnamefont {S.~S.}\ \bibnamefont {Parkin}},\ }\href@noop {} {\bibfield
  {journal} {\bibinfo  {journal} {Science Advances}\ }\textbf {\bibinfo
  {volume} {2}},\ \bibinfo {pages} {e1501870} (\bibinfo {year}
  {2016})}\BibitemShut {NoStop}%
\bibitem [{\citenamefont {K{\"u}bler}\ and\ \citenamefont
  {Felser}(2018)}]{kubler2018weyl}%
  \BibitemOpen
  \bibfield  {author} {\bibinfo {author} {\bibfnamefont {J.}~\bibnamefont
  {K{\"u}bler}}\ and\ \bibinfo {author} {\bibfnamefont {C.}~\bibnamefont
  {Felser}},\ }\href@noop {} {\bibfield  {journal} {\bibinfo  {journal} {EPL
  (Europhysics Letters)}\ }\textbf {\bibinfo {volume} {120}},\ \bibinfo {pages}
  {47002} (\bibinfo {year} {2018})}\BibitemShut {NoStop}%
\bibitem [{\citenamefont {Zhang}\ \emph {et~al.}(2017)\citenamefont {Zhang},
  \citenamefont {Sun}, \citenamefont {Yang}, \citenamefont {{\v{Z}}elezn{\`y}},
  \citenamefont {Parkin}, \citenamefont {Felser},\ and\ \citenamefont
  {Yan}}]{zhang2017strong}%
  \BibitemOpen
  \bibfield  {author} {\bibinfo {author} {\bibfnamefont {Y.}~\bibnamefont
  {Zhang}}, \bibinfo {author} {\bibfnamefont {Y.}~\bibnamefont {Sun}}, \bibinfo
  {author} {\bibfnamefont {H.}~\bibnamefont {Yang}}, \bibinfo {author}
  {\bibfnamefont {J.}~\bibnamefont {{\v{Z}}elezn{\`y}}}, \bibinfo {author}
  {\bibfnamefont {S.~P.}\ \bibnamefont {Parkin}}, \bibinfo {author}
  {\bibfnamefont {C.}~\bibnamefont {Felser}},\ and\ \bibinfo {author}
  {\bibfnamefont {B.}~\bibnamefont {Yan}},\ }\href@noop {} {\bibfield
  {journal} {\bibinfo  {journal} {Physical Review B}\ }\textbf {\bibinfo
  {volume} {95}},\ \bibinfo {pages} {075128} (\bibinfo {year}
  {2017})}\BibitemShut {NoStop}%
\bibitem [{\citenamefont {Mukherjee}\ \emph {et~al.}(2021)\citenamefont
  {Mukherjee}, \citenamefont {Suraj}, \citenamefont {Basumatary}, \citenamefont
  {Sethupathi},\ and\ \citenamefont {Raman}}]{mukherjee2020sign}%
  \BibitemOpen
  \bibfield  {author} {\bibinfo {author} {\bibfnamefont {J.}~\bibnamefont
  {Mukherjee}}, \bibinfo {author} {\bibfnamefont {T.~S.}\ \bibnamefont
  {Suraj}}, \bibinfo {author} {\bibfnamefont {H.}~\bibnamefont {Basumatary}},
  \bibinfo {author} {\bibfnamefont {K.}~\bibnamefont {Sethupathi}},\ and\
  \bibinfo {author} {\bibfnamefont {K.~V.}\ \bibnamefont {Raman}},\ }\href@noop
  {} {\bibfield  {journal} {\bibinfo  {journal} {Phys. Rev. Materials}\
  }\textbf {\bibinfo {volume} {5}},\ \bibinfo {pages} {014201} (\bibinfo {year}
  {2021})}\BibitemShut {NoStop}%
\bibitem [{\citenamefont {Kuroda}\ \emph {et~al.}(2017)\citenamefont {Kuroda},
  \citenamefont {Tomita}, \citenamefont {Suzuki}, \citenamefont {Bareille},
  \citenamefont {Nugroho}, \citenamefont {Goswami}, \citenamefont {Ochi},
  \citenamefont {Ikhlas}, \citenamefont {Nakayama}, \citenamefont {Akebi} \emph
  {et~al.}}]{kuroda2017evidence}%
  \BibitemOpen
  \bibfield  {author} {\bibinfo {author} {\bibfnamefont {K.}~\bibnamefont
  {Kuroda}}, \bibinfo {author} {\bibfnamefont {T.}~\bibnamefont {Tomita}},
  \bibinfo {author} {\bibfnamefont {M.-T.}\ \bibnamefont {Suzuki}}, \bibinfo
  {author} {\bibfnamefont {C.}~\bibnamefont {Bareille}}, \bibinfo {author}
  {\bibfnamefont {A.}~\bibnamefont {Nugroho}}, \bibinfo {author} {\bibfnamefont
  {P.}~\bibnamefont {Goswami}}, \bibinfo {author} {\bibfnamefont
  {M.}~\bibnamefont {Ochi}}, \bibinfo {author} {\bibfnamefont {M.}~\bibnamefont
  {Ikhlas}}, \bibinfo {author} {\bibfnamefont {M.}~\bibnamefont {Nakayama}},
  \bibinfo {author} {\bibfnamefont {S.}~\bibnamefont {Akebi}}, \emph {et~al.},\
  }\href@noop {} {\bibfield  {journal} {\bibinfo  {journal} {Nature Materials}\
  }\textbf {\bibinfo {volume} {16}},\ \bibinfo {pages} {1090} (\bibinfo {year}
  {2017})}\BibitemShut {NoStop}%
\bibitem [{\citenamefont {Yang}\ \emph {et~al.}(2017)\citenamefont {Yang},
  \citenamefont {Sun}, \citenamefont {Zhang}, \citenamefont {Shi},
  \citenamefont {Parkin},\ and\ \citenamefont {Yan}}]{yang2017topological}%
  \BibitemOpen
  \bibfield  {author} {\bibinfo {author} {\bibfnamefont {H.}~\bibnamefont
  {Yang}}, \bibinfo {author} {\bibfnamefont {Y.}~\bibnamefont {Sun}}, \bibinfo
  {author} {\bibfnamefont {Y.}~\bibnamefont {Zhang}}, \bibinfo {author}
  {\bibfnamefont {W.-J.}\ \bibnamefont {Shi}}, \bibinfo {author} {\bibfnamefont
  {S.~S.}\ \bibnamefont {Parkin}},\ and\ \bibinfo {author} {\bibfnamefont
  {B.}~\bibnamefont {Yan}},\ }\href@noop {} {\bibfield  {journal} {\bibinfo
  {journal} {New Journal of Physics}\ }\textbf {\bibinfo {volume} {19}},\
  \bibinfo {pages} {015008} (\bibinfo {year} {2017})}\BibitemShut {NoStop}%
\bibitem [{\citenamefont {Chen}\ \emph {et~al.}(2021)\citenamefont {Chen},
  \citenamefont {Tomita}, \citenamefont {Minami}, \citenamefont {Fu},
  \citenamefont {Koretsune}, \citenamefont {Kitatani}, \citenamefont
  {Muhammad}, \citenamefont {Nishio-Hamane}, \citenamefont {Ishii},
  \citenamefont {Ishii}, \citenamefont {Arita},\ and\ \citenamefont
  {Nakatsuji}}]{chen2021anomalous}%
  \BibitemOpen
  \bibfield  {author} {\bibinfo {author} {\bibfnamefont {T.}~\bibnamefont
  {Chen}}, \bibinfo {author} {\bibfnamefont {T.}~\bibnamefont {Tomita}},
  \bibinfo {author} {\bibfnamefont {S.}~\bibnamefont {Minami}}, \bibinfo
  {author} {\bibfnamefont {M.}~\bibnamefont {Fu}}, \bibinfo {author}
  {\bibfnamefont {T.}~\bibnamefont {Koretsune}}, \bibinfo {author}
  {\bibfnamefont {M.}~\bibnamefont {Kitatani}}, \bibinfo {author}
  {\bibfnamefont {I.}~\bibnamefont {Muhammad}}, \bibinfo {author}
  {\bibfnamefont {D.}~\bibnamefont {Nishio-Hamane}}, \bibinfo {author}
  {\bibfnamefont {R.}~\bibnamefont {Ishii}}, \bibinfo {author} {\bibfnamefont
  {F.}~\bibnamefont {Ishii}}, \bibinfo {author} {\bibfnamefont
  {R.}~\bibnamefont {Arita}},\ and\ \bibinfo {author} {\bibfnamefont
  {S.}~\bibnamefont {Nakatsuji}},\ }\href@noop {} {\bibfield  {journal}
  {\bibinfo  {journal} {Nature Communications}\ }\textbf {\bibinfo {volume}
  {12}},\ \bibinfo {pages} {572} (\bibinfo {year} {2021})}\BibitemShut
  {NoStop}%
\bibitem [{\citenamefont {Iwaki}\ \emph {et~al.}(2020)\citenamefont {Iwaki},
  \citenamefont {Kimata}, \citenamefont {Ikebuchi}, \citenamefont {Kobayashi},
  \citenamefont {Oda}, \citenamefont {Shiota}, \citenamefont {Ono},\ and\
  \citenamefont {Moriyama}}]{iwaki2020large}%
  \BibitemOpen
  \bibfield  {author} {\bibinfo {author} {\bibfnamefont {H.}~\bibnamefont
  {Iwaki}}, \bibinfo {author} {\bibfnamefont {M.}~\bibnamefont {Kimata}},
  \bibinfo {author} {\bibfnamefont {T.}~\bibnamefont {Ikebuchi}}, \bibinfo
  {author} {\bibfnamefont {Y.}~\bibnamefont {Kobayashi}}, \bibinfo {author}
  {\bibfnamefont {K.}~\bibnamefont {Oda}}, \bibinfo {author} {\bibfnamefont
  {Y.}~\bibnamefont {Shiota}}, \bibinfo {author} {\bibfnamefont
  {T.}~\bibnamefont {Ono}},\ and\ \bibinfo {author} {\bibfnamefont
  {T.}~\bibnamefont {Moriyama}},\ }\href@noop {} {\bibfield  {journal}
  {\bibinfo  {journal} {Applied Physics Letters}\ }\textbf {\bibinfo {volume}
  {116}},\ \bibinfo {pages} {022408} (\bibinfo {year} {2020})}\BibitemShut
  {NoStop}%
\bibitem [{\citenamefont {Liu}\ \emph {et~al.}(2017)\citenamefont {Liu},
  \citenamefont {Zhang}, \citenamefont {Liu}, \citenamefont {Ding},
  \citenamefont {Liu}, \citenamefont {Jafri}, \citenamefont {Hou},
  \citenamefont {Wang}, \citenamefont {Ma},\ and\ \citenamefont
  {Wu}}]{liu2017transition}%
  \BibitemOpen
  \bibfield  {author} {\bibinfo {author} {\bibfnamefont {Z.}~\bibnamefont
  {Liu}}, \bibinfo {author} {\bibfnamefont {Y.}~\bibnamefont {Zhang}}, \bibinfo
  {author} {\bibfnamefont {G.}~\bibnamefont {Liu}}, \bibinfo {author}
  {\bibfnamefont {B.}~\bibnamefont {Ding}}, \bibinfo {author} {\bibfnamefont
  {E.}~\bibnamefont {Liu}}, \bibinfo {author} {\bibfnamefont {H.~M.}\
  \bibnamefont {Jafri}}, \bibinfo {author} {\bibfnamefont {Z.}~\bibnamefont
  {Hou}}, \bibinfo {author} {\bibfnamefont {W.}~\bibnamefont {Wang}}, \bibinfo
  {author} {\bibfnamefont {X.}~\bibnamefont {Ma}},\ and\ \bibinfo {author}
  {\bibfnamefont {G.}~\bibnamefont {Wu}},\ }\href@noop {} {\bibfield  {journal}
  {\bibinfo  {journal} {Scientific reports}\ }\textbf {\bibinfo {volume} {7}},\
  \bibinfo {pages} {1} (\bibinfo {year} {2017})}\BibitemShut {NoStop}%
\bibitem [{\citenamefont {S{\"u}rgers}(2018)}]{surgers2018electrical}%
  \BibitemOpen
  \bibfield  {author} {\bibinfo {author} {\bibfnamefont {C.}~\bibnamefont
  {S{\"u}rgers}},\ }\href@noop {} {\bibfield  {journal} {\bibinfo  {journal}
  {Nature Electronics}\ }\textbf {\bibinfo {volume} {1}},\ \bibinfo {pages}
  {154} (\bibinfo {year} {2018})}\BibitemShut {NoStop}%
\bibitem [{\citenamefont {Liu}\ \emph {et~al.}(2018)\citenamefont {Liu},
  \citenamefont {Chen}, \citenamefont {Wang}, \citenamefont {Liu},
  \citenamefont {Wang}, \citenamefont {Feng}, \citenamefont {Yan},
  \citenamefont {Wang}, \citenamefont {Jiang}, \citenamefont {Coey} \emph
  {et~al.}}]{liu2018electrical}%
  \BibitemOpen
  \bibfield  {author} {\bibinfo {author} {\bibfnamefont {Z.}~\bibnamefont
  {Liu}}, \bibinfo {author} {\bibfnamefont {H.}~\bibnamefont {Chen}}, \bibinfo
  {author} {\bibfnamefont {J.}~\bibnamefont {Wang}}, \bibinfo {author}
  {\bibfnamefont {J.}~\bibnamefont {Liu}}, \bibinfo {author} {\bibfnamefont
  {K.}~\bibnamefont {Wang}}, \bibinfo {author} {\bibfnamefont {Z.}~\bibnamefont
  {Feng}}, \bibinfo {author} {\bibfnamefont {H.}~\bibnamefont {Yan}}, \bibinfo
  {author} {\bibfnamefont {X.}~\bibnamefont {Wang}}, \bibinfo {author}
  {\bibfnamefont {C.}~\bibnamefont {Jiang}}, \bibinfo {author} {\bibfnamefont
  {J.}~\bibnamefont {Coey}}, \emph {et~al.},\ }\href@noop {} {\bibfield
  {journal} {\bibinfo  {journal} {Nature Electronics}\ }\textbf {\bibinfo
  {volume} {1}},\ \bibinfo {pages} {172} (\bibinfo {year} {2018})}\BibitemShut
  {NoStop}%
\bibitem [{\citenamefont {Kimata}\ \emph {et~al.}(2019)\citenamefont {Kimata},
  \citenamefont {Chen}, \citenamefont {Kondou}, \citenamefont {Sugimoto},
  \citenamefont {Muduli}, \citenamefont {Ikhlas}, \citenamefont {Omori},
  \citenamefont {Tomita}, \citenamefont {MacDonald}, \citenamefont {Nakatsuji}
  \emph {et~al.}}]{kimata2019magnetic}%
  \BibitemOpen
  \bibfield  {author} {\bibinfo {author} {\bibfnamefont {M.}~\bibnamefont
  {Kimata}}, \bibinfo {author} {\bibfnamefont {H.}~\bibnamefont {Chen}},
  \bibinfo {author} {\bibfnamefont {K.}~\bibnamefont {Kondou}}, \bibinfo
  {author} {\bibfnamefont {S.}~\bibnamefont {Sugimoto}}, \bibinfo {author}
  {\bibfnamefont {P.~K.}\ \bibnamefont {Muduli}}, \bibinfo {author}
  {\bibfnamefont {M.}~\bibnamefont {Ikhlas}}, \bibinfo {author} {\bibfnamefont
  {Y.}~\bibnamefont {Omori}}, \bibinfo {author} {\bibfnamefont
  {T.}~\bibnamefont {Tomita}}, \bibinfo {author} {\bibfnamefont {A.~H.}\
  \bibnamefont {MacDonald}}, \bibinfo {author} {\bibfnamefont {S.}~\bibnamefont
  {Nakatsuji}}, \emph {et~al.},\ }\href@noop {} {\bibfield  {journal} {\bibinfo
   {journal} {Nature}\ }\textbf {\bibinfo {volume} {565}},\ \bibinfo {pages}
  {627} (\bibinfo {year} {2019})}\BibitemShut {NoStop}%
\bibitem [{\citenamefont {Zhang}\ \emph {et~al.}(2016)\citenamefont {Zhang},
  \citenamefont {Han}, \citenamefont {Yang}, \citenamefont {Sun}, \citenamefont
  {Zhang}, \citenamefont {Yan},\ and\ \citenamefont {Parkin}}]{zhang2016giant}%
  \BibitemOpen
  \bibfield  {author} {\bibinfo {author} {\bibfnamefont {W.}~\bibnamefont
  {Zhang}}, \bibinfo {author} {\bibfnamefont {W.}~\bibnamefont {Han}}, \bibinfo
  {author} {\bibfnamefont {S.-H.}\ \bibnamefont {Yang}}, \bibinfo {author}
  {\bibfnamefont {Y.}~\bibnamefont {Sun}}, \bibinfo {author} {\bibfnamefont
  {Y.}~\bibnamefont {Zhang}}, \bibinfo {author} {\bibfnamefont
  {B.}~\bibnamefont {Yan}},\ and\ \bibinfo {author} {\bibfnamefont {S.~S.}\
  \bibnamefont {Parkin}},\ }\href@noop {} {\bibfield  {journal} {\bibinfo
  {journal} {Science Advances}\ }\textbf {\bibinfo {volume} {2}},\ \bibinfo
  {pages} {e1600759} (\bibinfo {year} {2016})}\BibitemShut {NoStop}%
\bibitem [{\citenamefont {Ikhlas}\ \emph {et~al.}(2017)\citenamefont {Ikhlas},
  \citenamefont {Tomita}, \citenamefont {Koretsune}, \citenamefont {Suzuki},
  \citenamefont {Nishio-Hamane}, \citenamefont {Arita}, \citenamefont {Otani},\
  and\ \citenamefont {Nakatsuji}}]{ikhlas2017large}%
  \BibitemOpen
  \bibfield  {author} {\bibinfo {author} {\bibfnamefont {M.}~\bibnamefont
  {Ikhlas}}, \bibinfo {author} {\bibfnamefont {T.}~\bibnamefont {Tomita}},
  \bibinfo {author} {\bibfnamefont {T.}~\bibnamefont {Koretsune}}, \bibinfo
  {author} {\bibfnamefont {M.-T.}\ \bibnamefont {Suzuki}}, \bibinfo {author}
  {\bibfnamefont {D.}~\bibnamefont {Nishio-Hamane}}, \bibinfo {author}
  {\bibfnamefont {R.}~\bibnamefont {Arita}}, \bibinfo {author} {\bibfnamefont
  {Y.}~\bibnamefont {Otani}},\ and\ \bibinfo {author} {\bibfnamefont
  {S.}~\bibnamefont {Nakatsuji}},\ }\href@noop {} {\bibfield  {journal}
  {\bibinfo  {journal} {Nature Physics}\ }\textbf {\bibinfo {volume} {13}},\
  \bibinfo {pages} {1085} (\bibinfo {year} {2017})}\BibitemShut {NoStop}%
\bibitem [{\citenamefont {Hong}\ \emph {et~al.}(2020)\citenamefont {Hong},
  \citenamefont {Anand}, \citenamefont {Liu}, \citenamefont {Liu},
  \citenamefont {Arslan}, \citenamefont {Pearson}, \citenamefont
  {Bhattacharya},\ and\ \citenamefont {Jiang}}]{hong2020large}%
  \BibitemOpen
  \bibfield  {author} {\bibinfo {author} {\bibfnamefont {D.}~\bibnamefont
  {Hong}}, \bibinfo {author} {\bibfnamefont {N.}~\bibnamefont {Anand}},
  \bibinfo {author} {\bibfnamefont {C.}~\bibnamefont {Liu}}, \bibinfo {author}
  {\bibfnamefont {H.}~\bibnamefont {Liu}}, \bibinfo {author} {\bibfnamefont
  {I.}~\bibnamefont {Arslan}}, \bibinfo {author} {\bibfnamefont {J.~E.}\
  \bibnamefont {Pearson}}, \bibinfo {author} {\bibfnamefont {A.}~\bibnamefont
  {Bhattacharya}},\ and\ \bibinfo {author} {\bibfnamefont {J.}~\bibnamefont
  {Jiang}},\ }\href@noop {} {\bibfield  {journal} {\bibinfo  {journal}
  {Physical Review Materials}\ }\textbf {\bibinfo {volume} {4}},\ \bibinfo
  {pages} {094201} (\bibinfo {year} {2020})}\BibitemShut {NoStop}%
\bibitem [{\citenamefont {Wuttke}\ \emph {et~al.}(2019)\citenamefont {Wuttke},
  \citenamefont {Caglieris}, \citenamefont {Sykora}, \citenamefont
  {Scaravaggi}, \citenamefont {Wolter}, \citenamefont {Manna}, \citenamefont
  {S{\"u}ss}, \citenamefont {Shekhar}, \citenamefont {Felser}, \citenamefont
  {B{\"u}chner} \emph {et~al.}}]{wuttke2019berry}%
  \BibitemOpen
  \bibfield  {author} {\bibinfo {author} {\bibfnamefont {C.}~\bibnamefont
  {Wuttke}}, \bibinfo {author} {\bibfnamefont {F.}~\bibnamefont {Caglieris}},
  \bibinfo {author} {\bibfnamefont {S.}~\bibnamefont {Sykora}}, \bibinfo
  {author} {\bibfnamefont {F.}~\bibnamefont {Scaravaggi}}, \bibinfo {author}
  {\bibfnamefont {A.~U.}\ \bibnamefont {Wolter}}, \bibinfo {author}
  {\bibfnamefont {K.}~\bibnamefont {Manna}}, \bibinfo {author} {\bibfnamefont
  {V.}~\bibnamefont {S{\"u}ss}}, \bibinfo {author} {\bibfnamefont
  {C.}~\bibnamefont {Shekhar}}, \bibinfo {author} {\bibfnamefont
  {C.}~\bibnamefont {Felser}}, \bibinfo {author} {\bibfnamefont
  {B.}~\bibnamefont {B{\"u}chner}}, \emph {et~al.},\ }\href@noop {} {\bibfield
  {journal} {\bibinfo  {journal} {Physical Review B}\ }\textbf {\bibinfo
  {volume} {100}},\ \bibinfo {pages} {085111} (\bibinfo {year}
  {2019})}\BibitemShut {NoStop}%
\bibitem [{\citenamefont {Xu}\ \emph {et~al.}(2020)\citenamefont {Xu},
  \citenamefont {Li}, \citenamefont {Lu}, \citenamefont {Collignon},
  \citenamefont {Fu}, \citenamefont {Koo}, \citenamefont {Fauqu{\'e}},
  \citenamefont {Yan}, \citenamefont {Zhu},\ and\ \citenamefont
  {Behnia}}]{xu2020finite}%
  \BibitemOpen
  \bibfield  {author} {\bibinfo {author} {\bibfnamefont {L.}~\bibnamefont
  {Xu}}, \bibinfo {author} {\bibfnamefont {X.}~\bibnamefont {Li}}, \bibinfo
  {author} {\bibfnamefont {X.}~\bibnamefont {Lu}}, \bibinfo {author}
  {\bibfnamefont {C.}~\bibnamefont {Collignon}}, \bibinfo {author}
  {\bibfnamefont {H.}~\bibnamefont {Fu}}, \bibinfo {author} {\bibfnamefont
  {J.}~\bibnamefont {Koo}}, \bibinfo {author} {\bibfnamefont {B.}~\bibnamefont
  {Fauqu{\'e}}}, \bibinfo {author} {\bibfnamefont {B.}~\bibnamefont {Yan}},
  \bibinfo {author} {\bibfnamefont {Z.}~\bibnamefont {Zhu}},\ and\ \bibinfo
  {author} {\bibfnamefont {K.}~\bibnamefont {Behnia}},\ }\href@noop {}
  {\bibfield  {journal} {\bibinfo  {journal} {Science Advances}\ }\textbf
  {\bibinfo {volume} {6}},\ \bibinfo {pages} {eaaz3522} (\bibinfo {year}
  {2020})}\BibitemShut {NoStop}%
\bibitem [{\citenamefont {Wu}\ \emph {et~al.}(2020)\citenamefont {Wu},
  \citenamefont {Isshiki}, \citenamefont {Chen}, \citenamefont {Higo},
  \citenamefont {Nakatsuji},\ and\ \citenamefont {Otani}}]{wu2020magneto}%
  \BibitemOpen
  \bibfield  {author} {\bibinfo {author} {\bibfnamefont {M.}~\bibnamefont
  {Wu}}, \bibinfo {author} {\bibfnamefont {H.}~\bibnamefont {Isshiki}},
  \bibinfo {author} {\bibfnamefont {T.}~\bibnamefont {Chen}}, \bibinfo {author}
  {\bibfnamefont {T.}~\bibnamefont {Higo}}, \bibinfo {author} {\bibfnamefont
  {S.}~\bibnamefont {Nakatsuji}},\ and\ \bibinfo {author} {\bibfnamefont
  {Y.}~\bibnamefont {Otani}},\ }\href@noop {} {\bibfield  {journal} {\bibinfo
  {journal} {Applied Physics Letters}\ }\textbf {\bibinfo {volume} {116}},\
  \bibinfo {pages} {132408} (\bibinfo {year} {2020})}\BibitemShut {NoStop}%
\bibitem [{\citenamefont {Reichlova}\ \emph {et~al.}(2019)\citenamefont
  {Reichlova}, \citenamefont {Janda}, \citenamefont {Godinho}, \citenamefont
  {Markou}, \citenamefont {Kriegner}, \citenamefont {Schlitz}, \citenamefont
  {Zelezny}, \citenamefont {Soban}, \citenamefont {Bejarano}, \citenamefont
  {Schultheiss} \emph {et~al.}}]{reichlova2019imaging}%
  \BibitemOpen
  \bibfield  {author} {\bibinfo {author} {\bibfnamefont {H.}~\bibnamefont
  {Reichlova}}, \bibinfo {author} {\bibfnamefont {T.}~\bibnamefont {Janda}},
  \bibinfo {author} {\bibfnamefont {J.}~\bibnamefont {Godinho}}, \bibinfo
  {author} {\bibfnamefont {A.}~\bibnamefont {Markou}}, \bibinfo {author}
  {\bibfnamefont {D.}~\bibnamefont {Kriegner}}, \bibinfo {author}
  {\bibfnamefont {R.}~\bibnamefont {Schlitz}}, \bibinfo {author} {\bibfnamefont
  {J.}~\bibnamefont {Zelezny}}, \bibinfo {author} {\bibfnamefont
  {Z.}~\bibnamefont {Soban}}, \bibinfo {author} {\bibfnamefont
  {M.}~\bibnamefont {Bejarano}}, \bibinfo {author} {\bibfnamefont
  {H.}~\bibnamefont {Schultheiss}}, \emph {et~al.},\ }\href@noop {} {\bibfield
  {journal} {\bibinfo  {journal} {Nature communications}\ }\textbf {\bibinfo
  {volume} {10}},\ \bibinfo {pages} {1} (\bibinfo {year} {2019})}\BibitemShut
  {NoStop}%
\bibitem [{\citenamefont {Kiyohara}\ \emph {et~al.}(2016)\citenamefont
  {Kiyohara}, \citenamefont {Tomita},\ and\ \citenamefont
  {Nakatsuji}}]{kiyohara2016giant}%
  \BibitemOpen
  \bibfield  {author} {\bibinfo {author} {\bibfnamefont {N.}~\bibnamefont
  {Kiyohara}}, \bibinfo {author} {\bibfnamefont {T.}~\bibnamefont {Tomita}},\
  and\ \bibinfo {author} {\bibfnamefont {S.}~\bibnamefont {Nakatsuji}},\
  }\bibfield  {title} {\bibinfo {title} {Giant anomalous hall effect in the
  chiral antiferromagnet mn 3 ge},\ }\href@noop {} {\bibfield  {journal}
  {\bibinfo  {journal} {Physical Review Applied}\ }\textbf {\bibinfo {volume}
  {5}},\ \bibinfo {pages} {064009} (\bibinfo {year} {2016})}\BibitemShut
  {NoStop}%
\bibitem [{\citenamefont {Ohoyama}(1961)}]{ohoyama1961x}%
  \BibitemOpen
  \bibfield  {author} {\bibinfo {author} {\bibfnamefont {T.}~\bibnamefont
  {Ohoyama}},\ }\href@noop {} {\bibfield  {journal} {\bibinfo  {journal}
  {Journal of the Physical Society of Japan}\ }\textbf {\bibinfo {volume}
  {16}},\ \bibinfo {pages} {1995} (\bibinfo {year} {1961})}\BibitemShut
  {NoStop}%
\bibitem [{\citenamefont {Nagamiya}\ \emph {et~al.}(1982)\citenamefont
  {Nagamiya}, \citenamefont {Tomiyoshi},\ and\ \citenamefont
  {Yamaguchi}}]{nagamiya1982triangular}%
  \BibitemOpen
  \bibfield  {author} {\bibinfo {author} {\bibfnamefont {T.}~\bibnamefont
  {Nagamiya}}, \bibinfo {author} {\bibfnamefont {S.}~\bibnamefont
  {Tomiyoshi}},\ and\ \bibinfo {author} {\bibfnamefont {Y.}~\bibnamefont
  {Yamaguchi}},\ }\href@noop {} {\bibfield  {journal} {\bibinfo  {journal}
  {Solid State Communications}\ }\textbf {\bibinfo {volume} {42}},\ \bibinfo
  {pages} {385} (\bibinfo {year} {1982})}\BibitemShut {NoStop}%
\bibitem [{\citenamefont {Tomiyoshi}\ \emph {et~al.}(1983)\citenamefont
  {Tomiyoshi}, \citenamefont {Yamaguchi},\ and\ \citenamefont
  {Nagamiya}}]{tomiyoshi1983triangular}%
  \BibitemOpen
  \bibfield  {author} {\bibinfo {author} {\bibfnamefont {S.}~\bibnamefont
  {Tomiyoshi}}, \bibinfo {author} {\bibfnamefont {Y.}~\bibnamefont
  {Yamaguchi}},\ and\ \bibinfo {author} {\bibfnamefont {T.}~\bibnamefont
  {Nagamiya}},\ }\href@noop {} {\bibfield  {journal} {\bibinfo  {journal}
  {Journal of magnetism and magnetic materials}\ }\textbf {\bibinfo {volume}
  {31}},\ \bibinfo {pages} {629} (\bibinfo {year} {1983})}\BibitemShut
  {NoStop}%
\bibitem [{\citenamefont {Yamada}\ \emph {et~al.}(1988)\citenamefont {Yamada},
  \citenamefont {Sakai}, \citenamefont {Mori},\ and\ \citenamefont
  {Ohoyama}}]{yamada1988magnetic}%
  \BibitemOpen
  \bibfield  {author} {\bibinfo {author} {\bibfnamefont {N.}~\bibnamefont
  {Yamada}}, \bibinfo {author} {\bibfnamefont {H.}~\bibnamefont {Sakai}},
  \bibinfo {author} {\bibfnamefont {H.}~\bibnamefont {Mori}},\ and\ \bibinfo
  {author} {\bibfnamefont {T.}~\bibnamefont {Ohoyama}},\ }\href@noop {}
  {\bibfield  {journal} {\bibinfo  {journal} {Physica B+ C}\ }\textbf {\bibinfo
  {volume} {149}},\ \bibinfo {pages} {311} (\bibinfo {year}
  {1988})}\BibitemShut {NoStop}%
\bibitem [{\citenamefont {Brown}\ \emph {et~al.}(1990)\citenamefont {Brown},
  \citenamefont {Nunez}, \citenamefont {Tasset}, \citenamefont {Forsyth},\ and\
  \citenamefont {Radhakrishna}}]{brown1990determination}%
  \BibitemOpen
  \bibfield  {author} {\bibinfo {author} {\bibfnamefont {P.}~\bibnamefont
  {Brown}}, \bibinfo {author} {\bibfnamefont {V.}~\bibnamefont {Nunez}},
  \bibinfo {author} {\bibfnamefont {F.}~\bibnamefont {Tasset}}, \bibinfo
  {author} {\bibfnamefont {J.}~\bibnamefont {Forsyth}},\ and\ \bibinfo {author}
  {\bibfnamefont {P.}~\bibnamefont {Radhakrishna}},\ }\href@noop {} {\bibfield
  {journal} {\bibinfo  {journal} {Journal of Physics: Condensed Matter}\
  }\textbf {\bibinfo {volume} {2}},\ \bibinfo {pages} {9409} (\bibinfo {year}
  {1990})}\BibitemShut {NoStop}%
\bibitem [{\citenamefont {Qian}\ \emph {et~al.}(2014)\citenamefont {Qian},
  \citenamefont {Nayak}, \citenamefont {Kreiner}, \citenamefont {Schnelle},\
  and\ \citenamefont {Felser}}]{qian2014exchange}%
  \BibitemOpen
  \bibfield  {author} {\bibinfo {author} {\bibfnamefont {J.}~\bibnamefont
  {Qian}}, \bibinfo {author} {\bibfnamefont {A.}~\bibnamefont {Nayak}},
  \bibinfo {author} {\bibfnamefont {G.}~\bibnamefont {Kreiner}}, \bibinfo
  {author} {\bibfnamefont {W.}~\bibnamefont {Schnelle}},\ and\ \bibinfo
  {author} {\bibfnamefont {C.}~\bibnamefont {Felser}},\ }\href@noop {}
  {\bibfield  {journal} {\bibinfo  {journal} {Journal of Physics D: Applied
  Physics}\ }\textbf {\bibinfo {volume} {47}},\ \bibinfo {pages} {305001}
  (\bibinfo {year} {2014})}\BibitemShut {NoStop}%
\bibitem [{\citenamefont {Fang}\ \emph {et~al.}(2003)\citenamefont {Fang},
  \citenamefont {Nagaosa}, \citenamefont {Takahashi}, \citenamefont {Asamitsu},
  \citenamefont {Mathieu}, \citenamefont {Ogasawara}, \citenamefont {Yamada},
  \citenamefont {Kawasaki}, \citenamefont {Tokura},\ and\ \citenamefont
  {Terakura}}]{fang2003anomalous}%
  \BibitemOpen
  \bibfield  {author} {\bibinfo {author} {\bibfnamefont {Z.}~\bibnamefont
  {Fang}}, \bibinfo {author} {\bibfnamefont {N.}~\bibnamefont {Nagaosa}},
  \bibinfo {author} {\bibfnamefont {K.~S.}\ \bibnamefont {Takahashi}}, \bibinfo
  {author} {\bibfnamefont {A.}~\bibnamefont {Asamitsu}}, \bibinfo {author}
  {\bibfnamefont {R.}~\bibnamefont {Mathieu}}, \bibinfo {author} {\bibfnamefont
  {T.}~\bibnamefont {Ogasawara}}, \bibinfo {author} {\bibfnamefont
  {H.}~\bibnamefont {Yamada}}, \bibinfo {author} {\bibfnamefont
  {M.}~\bibnamefont {Kawasaki}}, \bibinfo {author} {\bibfnamefont
  {Y.}~\bibnamefont {Tokura}},\ and\ \bibinfo {author} {\bibfnamefont
  {K.}~\bibnamefont {Terakura}},\ }\href@noop {} {\bibfield  {journal}
  {\bibinfo  {journal} {Science}\ }\textbf {\bibinfo {volume} {302}},\ \bibinfo
  {pages} {92} (\bibinfo {year} {2003})}\BibitemShut {NoStop}%
\bibitem [{\citenamefont {{\v{S}}mejkal}\ \emph {et~al.}(2017)\citenamefont
  {{\v{S}}mejkal}, \citenamefont {Jungwirth},\ and\ \citenamefont
  {Sinova}}]{vsmejkal2017route}%
  \BibitemOpen
  \bibfield  {author} {\bibinfo {author} {\bibfnamefont {L.}~\bibnamefont
  {{\v{S}}mejkal}}, \bibinfo {author} {\bibfnamefont {T.}~\bibnamefont
  {Jungwirth}},\ and\ \bibinfo {author} {\bibfnamefont {J.}~\bibnamefont
  {Sinova}},\ }\href@noop {} {\bibfield  {journal} {\bibinfo  {journal}
  {physica status solidi (RRL)--Rapid Research Letters}\ }\textbf {\bibinfo
  {volume} {11}},\ \bibinfo {pages} {1700044} (\bibinfo {year}
  {2017})}\BibitemShut {NoStop}%
\bibitem [{\citenamefont {Duan}\ \emph {et~al.}(2015)\citenamefont {Duan},
  \citenamefont {Ren}, \citenamefont {Liu}, \citenamefont {Li}, \citenamefont
  {Liu},\ and\ \citenamefont {Zhang}}]{duan2015magnetic}%
  \BibitemOpen
  \bibfield  {author} {\bibinfo {author} {\bibfnamefont {T.}~\bibnamefont
  {Duan}}, \bibinfo {author} {\bibfnamefont {W.}~\bibnamefont {Ren}}, \bibinfo
  {author} {\bibfnamefont {W.}~\bibnamefont {Liu}}, \bibinfo {author}
  {\bibfnamefont {S.}~\bibnamefont {Li}}, \bibinfo {author} {\bibfnamefont
  {W.}~\bibnamefont {Liu}},\ and\ \bibinfo {author} {\bibfnamefont
  {Z.}~\bibnamefont {Zhang}},\ }\href@noop {} {\bibfield  {journal} {\bibinfo
  {journal} {Applied Physics Letters}\ }\textbf {\bibinfo {volume} {107}},\
  \bibinfo {pages} {082403} (\bibinfo {year} {2015})}\BibitemShut {NoStop}%
\bibitem [{\citenamefont {Jenkins}\ \emph {et~al.}(2019)\citenamefont
  {Jenkins}, \citenamefont {Chantrell}, \citenamefont {Klemmer},\ and\
  \citenamefont {Evans}}]{jenkins2019magnetic}%
  \BibitemOpen
  \bibfield  {author} {\bibinfo {author} {\bibfnamefont {S.}~\bibnamefont
  {Jenkins}}, \bibinfo {author} {\bibfnamefont {R.~W.}\ \bibnamefont
  {Chantrell}}, \bibinfo {author} {\bibfnamefont {T.~J.}\ \bibnamefont
  {Klemmer}},\ and\ \bibinfo {author} {\bibfnamefont {R.~F.}\ \bibnamefont
  {Evans}},\ }\href@noop {} {\bibfield  {journal} {\bibinfo  {journal}
  {Physical Review B}\ }\textbf {\bibinfo {volume} {100}},\ \bibinfo {pages}
  {220405} (\bibinfo {year} {2019})}\BibitemShut {NoStop}%
\bibitem [{\citenamefont {Vallejo-Fernandez}\ \emph {et~al.}(2007)\citenamefont
  {Vallejo-Fernandez}, \citenamefont {Fernandez-Outon},\ and\ \citenamefont
  {O’Grady}}]{vallejo2007measurement}%
  \BibitemOpen
  \bibfield  {author} {\bibinfo {author} {\bibfnamefont {G.}~\bibnamefont
  {Vallejo-Fernandez}}, \bibinfo {author} {\bibfnamefont {L.}~\bibnamefont
  {Fernandez-Outon}},\ and\ \bibinfo {author} {\bibfnamefont {K.}~\bibnamefont
  {O’Grady}},\ }\href@noop {} {\bibfield  {journal} {\bibinfo  {journal}
  {Applied Physics Letters}\ }\textbf {\bibinfo {volume} {91}},\ \bibinfo
  {pages} {212503} (\bibinfo {year} {2007})}\BibitemShut {NoStop}%
\bibitem [{\citenamefont {Chen}\ \emph
  {et~al.}(2020{\natexlab{a}})\citenamefont {Chen}, \citenamefont {Wang},
  \citenamefont {Xiao}, \citenamefont {Guo}, \citenamefont {Niu},\ and\
  \citenamefont {MacDonald}}]{chen2020manipulating}%
  \BibitemOpen
  \bibfield  {author} {\bibinfo {author} {\bibfnamefont {H.}~\bibnamefont
  {Chen}}, \bibinfo {author} {\bibfnamefont {T.-C.}\ \bibnamefont {Wang}},
  \bibinfo {author} {\bibfnamefont {D.}~\bibnamefont {Xiao}}, \bibinfo {author}
  {\bibfnamefont {G.-Y.}\ \bibnamefont {Guo}}, \bibinfo {author} {\bibfnamefont
  {Q.}~\bibnamefont {Niu}},\ and\ \bibinfo {author} {\bibfnamefont {A.~H.}\
  \bibnamefont {MacDonald}},\ }\href@noop {} {\bibfield  {journal} {\bibinfo
  {journal} {Physical Review B}\ }\textbf {\bibinfo {volume} {101}},\ \bibinfo
  {pages} {104418} (\bibinfo {year} {2020}{\natexlab{a}})}\BibitemShut
  {NoStop}%
\bibitem [{\citenamefont {Szunyogh}\ \emph {et~al.}(2009)\citenamefont
  {Szunyogh}, \citenamefont {Lazarovits}, \citenamefont {Udvardi},
  \citenamefont {Jackson},\ and\ \citenamefont {Nowak}}]{szunyogh2009giant}%
  \BibitemOpen
  \bibfield  {author} {\bibinfo {author} {\bibfnamefont {L.}~\bibnamefont
  {Szunyogh}}, \bibinfo {author} {\bibfnamefont {B.}~\bibnamefont
  {Lazarovits}}, \bibinfo {author} {\bibfnamefont {L.}~\bibnamefont {Udvardi}},
  \bibinfo {author} {\bibfnamefont {J.}~\bibnamefont {Jackson}},\ and\ \bibinfo
  {author} {\bibfnamefont {U.}~\bibnamefont {Nowak}},\ }\href@noop {}
  {\bibfield  {journal} {\bibinfo  {journal} {Physical Review B}\ }\textbf
  {\bibinfo {volume} {79}},\ \bibinfo {pages} {020403} (\bibinfo {year}
  {2009})}\BibitemShut {NoStop}%
\bibitem [{\citenamefont {Cable}\ \emph {et~al.}(1993)\citenamefont {Cable},
  \citenamefont {Wakabayashi},\ and\ \citenamefont
  {Radhakrishna}}]{cable1993magnetic}%
  \BibitemOpen
  \bibfield  {author} {\bibinfo {author} {\bibfnamefont {J.}~\bibnamefont
  {Cable}}, \bibinfo {author} {\bibfnamefont {N.}~\bibnamefont {Wakabayashi}},\
  and\ \bibinfo {author} {\bibfnamefont {P.}~\bibnamefont {Radhakrishna}},\
  }\href@noop {} {\bibfield  {journal} {\bibinfo  {journal} {Physical Review
  B}\ }\textbf {\bibinfo {volume} {48}},\ \bibinfo {pages} {6159} (\bibinfo
  {year} {1993})}\BibitemShut {NoStop}%
\bibitem [{\citenamefont {Manna}\ \emph {et~al.}(2018)\citenamefont {Manna},
  \citenamefont {Sun}, \citenamefont {Muechler}, \citenamefont {K{\"u}bler},\
  and\ \citenamefont {Felser}}]{manna2018heusler}%
  \BibitemOpen
  \bibfield  {author} {\bibinfo {author} {\bibfnamefont {K.}~\bibnamefont
  {Manna}}, \bibinfo {author} {\bibfnamefont {Y.}~\bibnamefont {Sun}}, \bibinfo
  {author} {\bibfnamefont {L.}~\bibnamefont {Muechler}}, \bibinfo {author}
  {\bibfnamefont {J.}~\bibnamefont {K{\"u}bler}},\ and\ \bibinfo {author}
  {\bibfnamefont {C.}~\bibnamefont {Felser}},\ }\href@noop {} {\bibfield
  {journal} {\bibinfo  {journal} {Nature Reviews Materials}\ }\textbf {\bibinfo
  {volume} {3}},\ \bibinfo {pages} {244} (\bibinfo {year} {2018})}\BibitemShut
  {NoStop}%
\bibitem [{\citenamefont {Ny{\'a}ri}\ \emph {et~al.}(2019)\citenamefont
  {Ny{\'a}ri}, \citenamefont {De{\'a}k},\ and\ \citenamefont
  {Szunyogh}}]{nyari2019weak}%
  \BibitemOpen
  \bibfield  {author} {\bibinfo {author} {\bibfnamefont {B.}~\bibnamefont
  {Ny{\'a}ri}}, \bibinfo {author} {\bibfnamefont {A.}~\bibnamefont
  {De{\'a}k}},\ and\ \bibinfo {author} {\bibfnamefont {L.}~\bibnamefont
  {Szunyogh}},\ }\href@noop {} {\bibfield  {journal} {\bibinfo  {journal}
  {Physical Review B}\ }\textbf {\bibinfo {volume} {100}},\ \bibinfo {pages}
  {144412} (\bibinfo {year} {2019})}\BibitemShut {NoStop}%
\bibitem [{\citenamefont {Liu}\ and\ \citenamefont
  {Balents}(2017)}]{liu2017anomalous}%
  \BibitemOpen
  \bibfield  {author} {\bibinfo {author} {\bibfnamefont {J.}~\bibnamefont
  {Liu}}\ and\ \bibinfo {author} {\bibfnamefont {L.}~\bibnamefont {Balents}},\
  }\href@noop {} {\bibfield  {journal} {\bibinfo  {journal} {Physical Review
  Letters}\ }\textbf {\bibinfo {volume} {119}},\ \bibinfo {pages} {087202}
  (\bibinfo {year} {2017})}\BibitemShut {NoStop}%
\bibitem [{\citenamefont {Soh}\ \emph {et~al.}(2020)\citenamefont {Soh},
  \citenamefont {de~Juan}, \citenamefont {Qureshi}, \citenamefont {Jacobsen},
  \citenamefont {Wang}, \citenamefont {Guo},\ and\ \citenamefont
  {Boothroyd}}]{soh2020ground}%
  \BibitemOpen
  \bibfield  {author} {\bibinfo {author} {\bibfnamefont {J.-R.}\ \bibnamefont
  {Soh}}, \bibinfo {author} {\bibfnamefont {F.}~\bibnamefont {de~Juan}},
  \bibinfo {author} {\bibfnamefont {N.}~\bibnamefont {Qureshi}}, \bibinfo
  {author} {\bibfnamefont {H.}~\bibnamefont {Jacobsen}}, \bibinfo {author}
  {\bibfnamefont {H.-Y.}\ \bibnamefont {Wang}}, \bibinfo {author}
  {\bibfnamefont {Y.-F.}\ \bibnamefont {Guo}},\ and\ \bibinfo {author}
  {\bibfnamefont {A.}~\bibnamefont {Boothroyd}},\ }\href@noop {} {\bibfield
  {journal} {\bibinfo  {journal} {Physical Review B}\ }\textbf {\bibinfo
  {volume} {101}},\ \bibinfo {pages} {140411} (\bibinfo {year}
  {2020})}\BibitemShut {NoStop}%
\bibitem [{\citenamefont {Chen}\ \emph
  {et~al.}(2020{\natexlab{b}})\citenamefont {Chen}, \citenamefont {Gaudet},
  \citenamefont {Dasgupta}, \citenamefont {Marcus}, \citenamefont {Lin},
  \citenamefont {Chen}, \citenamefont {Tomita}, \citenamefont {Ikhlas},
  \citenamefont {Zhao}, \citenamefont {Chen} \emph
  {et~al.}}]{chen2020antichiral}%
  \BibitemOpen
  \bibfield  {author} {\bibinfo {author} {\bibfnamefont {Y.}~\bibnamefont
  {Chen}}, \bibinfo {author} {\bibfnamefont {J.}~\bibnamefont {Gaudet}},
  \bibinfo {author} {\bibfnamefont {S.}~\bibnamefont {Dasgupta}}, \bibinfo
  {author} {\bibfnamefont {G.}~\bibnamefont {Marcus}}, \bibinfo {author}
  {\bibfnamefont {J.}~\bibnamefont {Lin}}, \bibinfo {author} {\bibfnamefont
  {T.}~\bibnamefont {Chen}}, \bibinfo {author} {\bibfnamefont {T.}~\bibnamefont
  {Tomita}}, \bibinfo {author} {\bibfnamefont {M.}~\bibnamefont {Ikhlas}},
  \bibinfo {author} {\bibfnamefont {Y.}~\bibnamefont {Zhao}}, \bibinfo {author}
  {\bibfnamefont {W.}~\bibnamefont {Chen}}, \emph {et~al.},\ }\href@noop {}
  {\bibfield  {journal} {\bibinfo  {journal} {Physical Review B}\ }\textbf
  {\bibinfo {volume} {102}},\ \bibinfo {pages} {054403} (\bibinfo {year}
  {2020}{\natexlab{b}})}\BibitemShut {NoStop}%
\bibitem [{\citenamefont {Xiao}\ \emph {et~al.}(2006)\citenamefont {Xiao},
  \citenamefont {Hu}, \citenamefont {Almasan}, \citenamefont {Sayles},\ and\
  \citenamefont {Maple}}]{xiao2006angular}%
  \BibitemOpen
  \bibfield  {author} {\bibinfo {author} {\bibfnamefont {H.}~\bibnamefont
  {Xiao}}, \bibinfo {author} {\bibfnamefont {T.}~\bibnamefont {Hu}}, \bibinfo
  {author} {\bibfnamefont {C.}~\bibnamefont {Almasan}}, \bibinfo {author}
  {\bibfnamefont {T.}~\bibnamefont {Sayles}},\ and\ \bibinfo {author}
  {\bibfnamefont {M.}~\bibnamefont {Maple}},\ }\href@noop {} {\bibfield
  {journal} {\bibinfo  {journal} {Physical Review B}\ }\textbf {\bibinfo
  {volume} {73}},\ \bibinfo {pages} {184511} (\bibinfo {year}
  {2006})}\BibitemShut {NoStop}%
\bibitem [{\citenamefont {Hu}\ \emph {et~al.}(2012)\citenamefont {Hu},
  \citenamefont {Xiao}, \citenamefont {Gyawali}, \citenamefont {Wen},\ and\
  \citenamefont {Almasan}}]{hu2012superconductivity}%
  \BibitemOpen
  \bibfield  {author} {\bibinfo {author} {\bibfnamefont {T.}~\bibnamefont
  {Hu}}, \bibinfo {author} {\bibfnamefont {H.}~\bibnamefont {Xiao}}, \bibinfo
  {author} {\bibfnamefont {P.}~\bibnamefont {Gyawali}}, \bibinfo {author}
  {\bibfnamefont {H.}~\bibnamefont {Wen}},\ and\ \bibinfo {author}
  {\bibfnamefont {C.}~\bibnamefont {Almasan}},\ }\href@noop {} {\bibfield
  {journal} {\bibinfo  {journal} {Physical Review B}\ }\textbf {\bibinfo
  {volume} {85}},\ \bibinfo {pages} {134516} (\bibinfo {year}
  {2012})}\BibitemShut {NoStop}%
\bibitem [{\citenamefont {Tomiyoshi}\ and\ \citenamefont
  {Yamaguchi}(1982)}]{tomiyoshi1982magnetic}%
  \BibitemOpen
  \bibfield  {author} {\bibinfo {author} {\bibfnamefont {S.}~\bibnamefont
  {Tomiyoshi}}\ and\ \bibinfo {author} {\bibfnamefont {Y.}~\bibnamefont
  {Yamaguchi}},\ }\href@noop {} {\bibfield  {journal} {\bibinfo  {journal}
  {Journal of the Physical Society of Japan}\ }\textbf {\bibinfo {volume}
  {51}},\ \bibinfo {pages} {2478} (\bibinfo {year} {1982})}\BibitemShut
  {NoStop}%
\bibitem [{\citenamefont {Herak}\ \emph {et~al.}(2015)\citenamefont {Herak},
  \citenamefont {{\v{Z}}ili{\'c}}, \citenamefont {{\v{C}}alogovi{\'c}},\ and\
  \citenamefont {Berger}}]{herak2015torque}%
  \BibitemOpen
  \bibfield  {author} {\bibinfo {author} {\bibfnamefont {M.}~\bibnamefont
  {Herak}}, \bibinfo {author} {\bibfnamefont {D.}~\bibnamefont
  {{\v{Z}}ili{\'c}}}, \bibinfo {author} {\bibfnamefont {D.~M.}\ \bibnamefont
  {{\v{C}}alogovi{\'c}}},\ and\ \bibinfo {author} {\bibfnamefont
  {H.}~\bibnamefont {Berger}},\ }\href@noop {} {\bibfield  {journal} {\bibinfo
  {journal} {Physical Review B}\ }\textbf {\bibinfo {volume} {91}},\ \bibinfo
  {pages} {174436} (\bibinfo {year} {2015})}\BibitemShut {NoStop}%
\bibitem [{\citenamefont {Hong}\ \emph {et~al.}(2016)\citenamefont {Hong},
  \citenamefont {Jo}, \citenamefont {Choi}, \citenamefont {Lee}, \citenamefont
  {Choi},\ and\ \citenamefont {Kang}}]{hong2016large}%
  \BibitemOpen
  \bibfield  {author} {\bibinfo {author} {\bibfnamefont {Y.}~\bibnamefont
  {Hong}}, \bibinfo {author} {\bibfnamefont {Y.}~\bibnamefont {Jo}}, \bibinfo
  {author} {\bibfnamefont {H.~Y.}\ \bibnamefont {Choi}}, \bibinfo {author}
  {\bibfnamefont {N.}~\bibnamefont {Lee}}, \bibinfo {author} {\bibfnamefont
  {Y.~J.}\ \bibnamefont {Choi}},\ and\ \bibinfo {author} {\bibfnamefont
  {W.}~\bibnamefont {Kang}},\ }\href@noop {} {\bibfield  {journal} {\bibinfo
  {journal} {Physical Review B}\ }\textbf {\bibinfo {volume} {93}},\ \bibinfo
  {pages} {094406} (\bibinfo {year} {2016})}\BibitemShut {NoStop}%
\bibitem [{\citenamefont {Fujita}(2017)}]{fujita2017field}%
  \BibitemOpen
  \bibfield  {author} {\bibinfo {author} {\bibfnamefont {H.}~\bibnamefont
  {Fujita}},\ }\href@noop {} {\bibfield  {journal} {\bibinfo  {journal}
  {physica status solidi (RRL)--Rapid Research Letters}\ }\textbf {\bibinfo
  {volume} {11}},\ \bibinfo {pages} {1600360} (\bibinfo {year}
  {2017})}\BibitemShut {NoStop}%
\bibitem [{\citenamefont {Meiklejohn}\ and\ \citenamefont
  {Bean}(1957)}]{meiklejohn1957new}%
  \BibitemOpen
  \bibfield  {author} {\bibinfo {author} {\bibfnamefont {W.~H.}\ \bibnamefont
  {Meiklejohn}}\ and\ \bibinfo {author} {\bibfnamefont {C.~P.}\ \bibnamefont
  {Bean}},\ }\href@noop {} {\bibfield  {journal} {\bibinfo  {journal} {Physical
  Review}\ }\textbf {\bibinfo {volume} {105}},\ \bibinfo {pages} {904}
  (\bibinfo {year} {1957})}\BibitemShut {NoStop}%
\bibitem [{\citenamefont {Nogu\'es}\ and\ \citenamefont
  {Schuller}(1999)}]{nogues1999exchange}%
  \BibitemOpen
  \bibfield  {author} {\bibinfo {author} {\bibfnamefont {J.}~\bibnamefont
  {Nogu\'es}}\ and\ \bibinfo {author} {\bibfnamefont {I.~K.}\ \bibnamefont
  {Schuller}},\ }\href@noop {} {\bibfield  {journal} {\bibinfo  {journal}
  {Journal of Magnetism and Magnetic Materials}\ }\textbf {\bibinfo {volume}
  {192}},\ \bibinfo {pages} {203} (\bibinfo {year} {1999})}\BibitemShut
  {NoStop}%
\bibitem [{\citenamefont {Gomonay}\ and\ \citenamefont
  {Loktev}(2005)}]{gomonay2005mechanism}%
  \BibitemOpen
  \bibfield  {author} {\bibinfo {author} {\bibfnamefont {E.}~\bibnamefont
  {Gomonay}}\ and\ \bibinfo {author} {\bibfnamefont {V.}~\bibnamefont
  {Loktev}},\ }\href@noop {} {\bibfield  {journal} {\bibinfo  {journal}
  {Physics of the Solid State}\ }\textbf {\bibinfo {volume} {47}},\ \bibinfo
  {pages} {1755} (\bibinfo {year} {2005})}\BibitemShut {NoStop}%
\bibitem [{\citenamefont {Gomonay}\ and\ \citenamefont
  {Loktev}(2007)}]{gomonay2007shape}%
  \BibitemOpen
  \bibfield  {author} {\bibinfo {author} {\bibfnamefont {H.~V.}\ \bibnamefont
  {Gomonay}}\ and\ \bibinfo {author} {\bibfnamefont {V.~M.}\ \bibnamefont
  {Loktev}},\ }\href@noop {} {\bibfield  {journal} {\bibinfo  {journal}
  {Physical Review B}\ }\textbf {\bibinfo {volume} {75}},\ \bibinfo {pages}
  {174439} (\bibinfo {year} {2007})}\BibitemShut {NoStop}%
\bibitem [{\citenamefont {Bauer}\ and\ \citenamefont
  {Rotter}(2009)}]{bauer2009mag}%
  \BibitemOpen
  \bibfield  {author} {\bibinfo {author} {\bibfnamefont {E.}~\bibnamefont
  {Bauer}}\ and\ \bibinfo {author} {\bibfnamefont {M.}~\bibnamefont {Rotter}},\
  }\bibfield  {title} {\bibinfo {title} {Magnetism of complex etallic alloys:
  Crystalline electric field effects},\ }in\ \href@noop {} {\emph {\bibinfo
  {booktitle} {Properties and Applications of Complex Intermetallics}}},\
  \bibinfo {editor} {edited by\ \bibinfo {editor} {\bibfnamefont
  {E.}~\bibnamefont {Belin-Ferr{\'e}}}}\ (\bibinfo  {publisher} {World
  Scientific},\ \bibinfo {address} {Singapore},\ \bibinfo {year} {2009})\ pp.\
  \bibinfo {pages} {183--248}\BibitemShut {NoStop}%
\end{thebibliography}

%

\end{document}